\documentclass[prd,onecolumn,notitlepage,eqsecnum,nofootinbib,floatfix]{revtex4-1}
\usepackage{amssymb}
\usepackage{amsmath}
\usepackage{amsfonts} 
\usepackage{bm}
\usepackage{graphicx}
\usepackage{comment} 
\DeclareSymbolFont{EulerScript}{U}{eus}{m}{n}
\SetSymbolFont{EulerScript}{bold}{U}{eus}{b}{n}
\DeclareSymbolFontAlphabet\scrpt{EulerScript}
\newcommand{\Lie}{{\pounds}} 
\newcommand{\KK}{{\scrpt K}} 
\newcommand{\bkt}[2]{{\langle #1 | #2 \rangle}} 
\newcommand{\stf}[1]{{\langle #1 \rangle}}
\newcommand{\gothp}{\mathfrak{p}} 
\newcommand{\gothj}{\mathfrak{j}} 
\newcommand{\gothn}{\mathfrak{n}} 
\newcommand{\gothm}{\mathfrak{m}}
\newcommand{\gothu}{\mathfrak{u}}
\newcommand{\psum}{\sideset{}{'}\sum}
\allowdisplaybreaks
\begin{document}
\title{Gravitomagnetic tidal resonances in neutron-star binary inspirals} 
\author{Eric Poisson}  
\affiliation{Department of Physics, University of Guelph, Guelph,
  Ontario, N1G 2W1, Canada} 
\date{April 28, 2020} 
\begin{abstract} 
A compact binary system implicating at least one rotating neutron star undergoes a sequence of four gravitomagnetic tidal resonances as it inspirals toward its final merger. These resonances have a dynamical impact on the binary's orbital motion, and thus on the phasing of the emitted gravitational waves. The resonances are produced by the inertial modes of vibration of the rotating star, and they occur when the orbital frequency becomes momentarily equal to a mode eigenfrequency. Four distinct modes are involved, and their eigenfrequencies are equal, up to a numerical factor of order unity, to the star's rotational angular velocity. The resonances occur within the frequency band of interferometric gravitational-wave detectors when the star spins at a frequency that lies within this band; the phenomenon is therefore of relevance to LIGO/Virgo for rotation rates comparable to $100\ \mbox{Hz}$. The resonances are driven by the gravitomagnetic tidal field created by the companion star; this is described by a post-Newtonian vector potential (the time-space components of the metric tensor), which is produced by the mass currents associated with the orbital motion. The gravitomagnetic tidal resonances were identified previously by Flanagan and Racine [Phys.\ Rev.\ D {\bf 75}, 044001 (2007)], but these authors accounted only for the response of a single mode, the $r$-mode, a special case of inertial modes. All four relevant inertial modes (including the $r$-mode) are included in the analysis presented in this paper. The total accumulated gravitational-wave phase shift caused by the gravitomagnetic tidal resonances is shown to range from approximately $10^{-2}$ radians when the spin and orbital angular momenta are aligned, to approximately $10^{-1}$ radians when the angular momenta are anti-aligned. Such phase shifts are small, but they will become measurable in the coming decades with the deployment of the next generation of gravitational-wave detectors (Cosmic Explorer, Einstein Telescope); they might even come to light within this decade, thanks to planned improvements in the current detectors (LIGO A+). With good constraints on the binary masses and spins gathered from the inspiral waveform, the phase shifts incurred during the gravitomagnetic tidal resonances deliver information regarding the internal structure of the rotating neutron star, and therefore on the equation of state of nuclear matter at high densities. 
\end{abstract} 

\maketitle

\section{Introduction and summary} 
\label{sec:intro} 

\subsection{Context: resonant excitation of stellar normal modes during binary inspirals} 

The near-simultaneous measurement of a neutron-star binary merger in gravitational waves (GW170817 \cite{GW170817:17}), gamma rays \cite{GW170817fermi:18, GW170817integral:17}, and other parts of the electromagnetic-wave spectrum \cite{GW170817multi:17} marked the beginning of a new era of multimessenger astronomy that promises, in the fullness of time, to reveal the intimate details of neutron-star interiors. This effort is reinforced with the recent deployment of the NICER instrument \cite{gendreau-etal:16} on board the International Space Station, which measures X-ray emissions from pulsars. Of primary interest is to elucidate the nature of nuclear matter at very high densities, which determines the stellar structure through the equation of state. The measurement of the mass and radius of neutron stars, through all conceivable means, provides constraints on the equation of state, and this in turn constrains the physics of nuclear matter at supernuclear densities. 

Flanagan and Hinderer \cite{flanagan-hinderer:08} recognized that details of the neutron-star interior could be revealed by gravitational-wave measurements of the star's tidal polarizability, which varies sensitively with the equation of state. This observation initiated a large effort \cite{flanagan-hinderer:08, hinderer-etal:10, vines-flanagan-hinderer:11, baiotti-etal:11, pannarale-etal:11, ferrari-gualtieri-maselli:12, lackey-etal:12, damour-nagar-villain:12, maselli-etal:12, maselli-etal:13, vines-flanagan:13, read-etal:13, maselli-gualtieri-ferrari:13, favata:14, yagi-yunes:14, lackey-etal:14, wade-etal:14, bernuzzi-etal:15, lackey-etal:17, cullen-etal:17, harry-hinderer:18, piekarewicz-fattoyev:19} to determine just how well this measurement can be made, and what information it can deliver on the star's internal structure. Such a measurement was attempted with GW170817 \cite{GW170817:18}, and it delivered an astrophysically interesting upper bound. On the other hand, the NICER measurement of the mass and radius of the pulsar PSR J0030+0451 has already provided useful constraints on the nuclear equation of state \cite{miller-etal:19}.

All means of measurement of neutron-star properties are subjected to systematic errors than can be difficult to quantify and control. It is therefore important to identify many independent ways of performing these measurements, in the hope that a consistent picture will emerge, in spite of the systematic errors. Other ways to gain access to the internal physics of neutron stars through gravitational waves include exploiting post-merger information at high frequencies \cite{bauswein-janka:12, hotokezaka-etal:13, takami-rezzolla-baiotti:14, bauswein-stergioulas:15, martynov-etal:19}, and accurate modeling of dynamical tides, which implicate the $f$-mode of vibration of neutron stars \cite{hinderer-etal:16, steinhoff-etal:16, schmidt-hinderer:19, ma-yu-chen:20}. A recent study by Pan {\it et al.}\ \cite{pan-etal:20} reveals that interface modes \cite{mcdermott-vanhorn-hansen:88, pan-etal:20}, which result from an interaction between the fluid core and solid crust of a neutron star, can be manifested during a binary inspiral, leading to phasing effects that can be measured by an interferometric detector in the LIGO A+ configuration \cite{LIGOAplus}. 

Yet another way, the focus of this paper, is to rely on the resonant excitation of a star's normal modes of vibration. The main idea is this: The tidal forces exerted by the neutron star's companion serve to perturb the star, and the perturbation can be decomposed into normal modes, which behave as a collection of driven harmonic oscillators. At some time during the inspiral, the frequency of the tidal field (either the orbital frequency or twice the orbital frequency) becomes momentarily equal to a mode's eigenfrequency, and this provokes the development of a resonance. The rapid growth of the mode during resonance takes energy away from the orbital motion\footnote{As we shall see below, in some circumstances the resonance can provide energy to the orbital motion, at the expense of the star's rotational energy.}, and this affects the inspiral in a way that can be measured in the phasing of the emitted gravitational waves. 

Early studies of this phenomenon \cite{lai:94, reisenegger-goldreich:94, shibata:94, kokkotas-schafer:95, lai:97, ho-lai:99} concluded that $f$-modes cannot become resonant (because the mode frequency is too large, beyond the LIGO/Virgo frequency band), and that while $g$-modes can be resonantly excited (because their frequencies are sufficiently low), they lead to orbital changes that are too small to be detected (because the coupling between tidal forces and $g$-modes is very small). The interface modes studied by Pan {\it et al.}\ \cite{pan-etal:20} do not typically achieve resonance. 

Rotating neutron stars offer a wider spectrum of normal modes, and therefore provide a larger set of possibilities for resonant tidal interactions. The focus of this paper is with the {\it inertial modes} first identified by Lockitch and Friedman \cite{lockitch-friedman:99}  (see also Refs.~\cite{yoshida-lee:00, wu:05, passamonti-etal:09} for additional studies). These modes are predominantly perturbations of the star's velocity field, governed by a restoring force supplied by the stellar rotation (the Coriolis force, when viewed in the star's corotating frame). The property that makes these modes interesting for resonant tidal interactions is that their frequencies are of the same order of magnitude as the star's rotational angular velocity. For a star spinning with a frequency comparable to $100\ \mbox{Hz}$, around the peak frequency of the LIGO/Virgo interferometers, the inertial modes permit the development of resonances that have a measurable impact on the phasing of the gravitational waves. The inertial modes include $r$-modes \cite{papaloizou-pringle:78, provost-berthomieu-rocca:81, saio:82, smeyers-martens:83, lindblom-mendell-owen:99} as special members; while an $r$-mode gives rise to a velocity perturbation of pure axial parity, the decomposition of a generic inertial mode in vector harmonics involves both polar and axial terms.   

The coupling of inertial modes to a Newtonian tidal field was first investigated by Lai and Wu \cite{lai-wu:06} (see also Ref.~\cite{xu-lai:17} for a recent refinement that also includes $g$-modes), who concluded that extremely high rotational frequencies (beyond $100\ \mbox{Hz}$) would be required to achieve a measurable phase shift in the gravitational-wave signal. Flanagan and Racine \cite{flanagan-racine:07}, however, pointed out that inertial modes couple more strongly to a post-Newtonian tidal field. While the Newtonian field is associated with second derivatives of the gravitational potential $U$, generated by the mass density of the orbiting companion, the post-Newtonian field is given by second derivatives of a vector potential $U_j$, produced by the mass currents associated with the companion's orbital motion. Drawing an analogy with the vector potential of Maxwell's electrodynamics, $U_j$ is often called a gravitomagnetic potential, and the resulting tidal field described as a gravitomagnetic field. The stronger coupling of inertial modes with a gravitomagnetic tidal field led Flanagan and Racine to a more optimistic conclusion regarding the measurability of the resonant interaction.  

\subsection{This work: gravitomagnetic tidal resonances from inertial modes} 

Flanagan and Racine \cite{flanagan-racine:07} calculated the dynamical impact on a binary inspiral, and on the phasing of the emitted gravitational waves, of a resonant gravitomagnetic tidal interaction. In their setup, the tidal field couples to an inertial mode of the rotating neutron star, and a resonance occurs when the orbital frequency momentarily becomes equal to the mode's eigenfrequency. During resonance the velocity perturbation within the star gives rise to a nonvanishing mass-current quadrupole moment\footnote{This is essentially the integral of the current density $\rho\, \delta \bm{v}$, where $\rho$ is the mass density and $\delta \bm{v}$ the velocity perturbation, multiplied by two powers of $\bm{x}$, the position vector relative to the star's center of mass.}  $J^{jk}$ which, in turn, produces another contribution to the gravitomagnetic vector potential that affects the binary's orbital motion. So while the coupling between orbit and mode is mediated by the gravitomagnetic tidal field created by the companion, the coupling between mode and orbit is mediated by the current quadrupole moment developed within the neutron star.  

Flanagan and Racine, however, examined only the coupling with $r$-modes, in spite of the fact that other inertial modes do participate in the tidal interaction. My purpose with this paper is to incorporate these additional modes in the analysis, and therefore to produce a more complete understanding of the resonant tidal interaction. As we shall see presently, proper inclusion of all relevant inertial modes produces a qualitatively different picture of the tidal encounter.

Inertial modes of a rotating star are labelled by two integers. The first is $m$, which determines the $e^{im\phi}$ dependence of the velocity perturbation on the azimuthal angle $\phi$; because a mode with a negative value of $m$ is the complex conjugate of a mode with the corresponding positive value, it is sufficient to take $m \geq 0$. The second integer is $n$, which sequences the infinity of overtones for each value of $m$. Each inertial mode comes with a distinct eigenfrequency $\omega_n^m = w_n^m \Omega$, where $\Omega$ is the star's rotational angular velocity, and 
$w_n^m$ is a number of order unity.

\begin{figure} 
\includegraphics[width=0.7\linewidth]{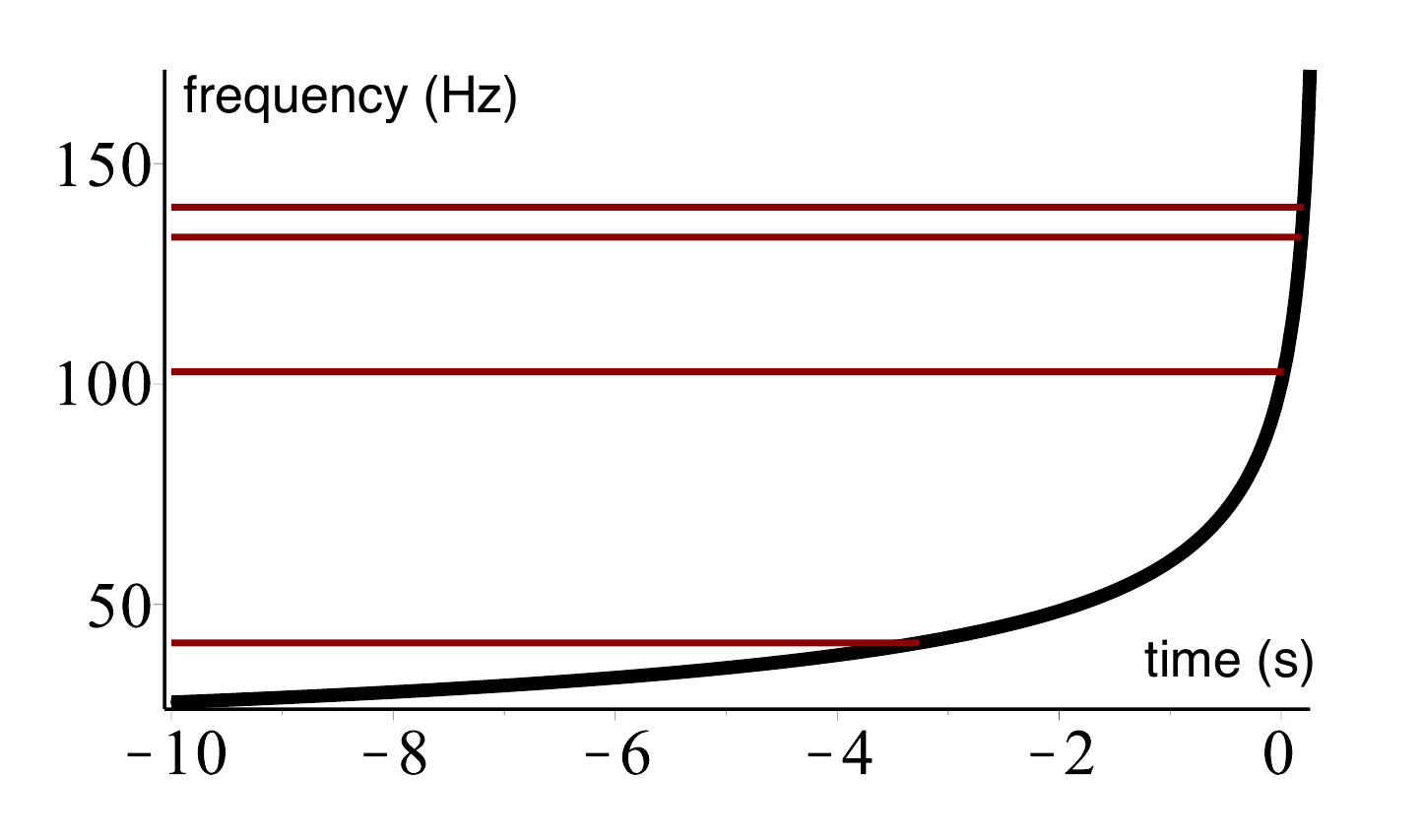}
\caption{Resonant tidal interaction during a binary inspiral. The thick black curve represents the sweep in orbital frequency that occurs during an inspiral. The binary parameters are given by the fiducial values adopted in Eq.~(\ref{DeltaPhi_intro}); in particular, the star's rotational frequency is chosen to be 100 Hz. The zero of time corresponds to a matching orbital frequency of 100 Hz. The thin horizontal lines mark the frequencies of the four inertial modes discussed in the text; a resonance takes place when the orbital frequency momentarily becomes equal to one of these frequencies. For the stellar model adopted in this plot --- Eq.~(\ref{rho_kmodel_intro}) with $k=1$ --- the lowest resonant frequency (41 Hz) corresponds to the mode labelled $(m = 1, n = \mbox{II})$, the second one (103 Hz) corresponds to $(m = 0, n = \mbox{I})$, the third (133 Hz) corresponds to the $r$-mode $(m=2, n = \bullet)$, and the highest frequency (140 Hz) corresponds to $(m=1, n=\mbox{I})$.}  
\label{fig:inspiral} 
\end{figure} 

I show in the technical sections of the paper that the resonant tidal interaction implicates four dominant modes, those that possess the strongest couplings with the gravitomagnetic tidal field; the situation is illustrated in Fig.~\ref{fig:inspiral}. The first is an $r$-mode with $m=2$, which is labelled $(m = 2, n = \bullet)$. The second is a $m=1$ mode with a positive frequency, to which I assign the label $(m = 1, n = \mbox{I})$. The third is another $m=1$ mode with a negative frequency, labelled $(m = 1, n = \mbox{II})$. The fourth is actually a complex-conjugate pair of $m = 0$ modes, with positive and negative frequencies of equal magnitude; this pair of modes is labelled $(m = 0, n = \mbox{I})$. Each mode gives rise to its own resonant tidal interaction, so that the encounter is actually a succession of four resonances; the total accumulated gravitational-wave phase shift is the sum of four contributions. I find that this phase shift is given by   
\begin{equation} 
\Delta\Phi_{\rm GW} = \gamma(\iota) \biggl( \frac{R}{10\ \mbox{km}} \biggr)^4 
\biggl( \frac{1.4\ M_\odot}{M} \biggr)^2 
\biggl( \frac{1.4\ M_\odot}{M'} \biggr) 
\biggl( \frac{2.8\ M_\odot}{M_{\rm tot}} \biggr)^{1/3} 
\biggl( \frac{\Omega/2\pi}{100\ \mbox{Hz}} \biggr)^{2/3}, 
\label{DeltaPhi_intro} 
\end{equation} 
where
\begin{equation}
\gamma(\iota) := \gothu_\bullet^2\, \sin^2\iota(\cos\iota + 1)^2 
+ \gothu_{\rm I}^1\, (\cos\iota + 1)^2 (2\cos\iota - 1)^2 
+ \gothu_{\rm II}^1\, (\cos\iota - 1)^2 (2\cos\iota + 1)^2
+ \gothu_{\rm I}^0\, \sin^2\iota \cos^2\iota, 
\label{gamma_intro} 
\end{equation} 
and where each number $\gothu_n^m$, defined by Eq.~(\ref{unm_def}) in the main text, is essentially a normalized overlap integral between the mode $(m , n)$ and the gravitomagnetic tidal field.  

In Eq.~(\ref{DeltaPhi_intro}), $R$ is the radius of the rotating neutron star, normalized by a fiducial value of $10\ \mbox{km}$, $M$ is its mass, normalized by a fiducial value of $1.4\ M_\odot$, and $\Omega$ is the rotational angular velocity, normalized with a corresponding frequency of $100\ \mbox{Hz}$; the second mass $M'$, also normalized with $1.4\ M_\odot$, is the companion's mass, and $M_{\rm tot} = M + M'$ is the binary's total mass. In Eq.~(\ref{gamma_intro}), $\iota$ is the orbit's inclination, the angle between the star's spin vector and the orbital angular-momentum vector. The numbers $\gothu_n^m$ are listed in Table~\ref{tab:uvalues_intro} for stellar models defined by the density function
\begin{equation}
\rho = C \biggl[ \frac{\sin(\pi r/R)}{(\pi r/R)} \biggr]^k,
\label{rho_kmodel_intro} 
\end{equation}
where $r$ is the radial coordinate inside the star, and $C$, $k$ are constants; low values of $k$ correspond to stars of nearly uniform density, while high values correspond to centrally dense stars. The model with $k=1$ results from the polytropic equation of state $p = K\rho^2$, where $p$ is the pressure and $K$ a constant. This makes a crude but serviceable model for a realistic equation of state for neutron-star matter. The stellar model with $k=1$ can therefore be considered to be the most realistic one.   

\begin{table} 
\caption{\label{tab:uvalues_intro} Numerical results for the coefficients $\gothu_n^m$, defined by Eq.~(\ref{unm_def}) in the main text, obtained for the density models of Eq.~(\ref{rho_kmodel_intro}). The first column gives $\gamma$ for $\iota = \pi$, while the second column lists $\gamma$ for $\iota = 0$; these are defined in Eq.~(\ref{gamma_intro}).} 
\begin{ruledtabular} 
\begin{tabular}{ccccccc}
$k$ & $\gamma(\pi)$ & $\gamma(0)$ & $\gothu_\bullet^2$ & $\gothu_{\rm I}^1$ & $\gothu_{\rm II}^1$ & $\gothu_{\rm I}^0$ \\ 
\hline
  0 & 2.3141e-1 & 4.7264e-2 & -6.1322e-2 & 1.1816e-2 & 5.7853e-2 & 2.8195e-1 \\
  1 & 1.4877e-1 & 4.1400e-2 & -2.9403e-2 & 1.0350e-2 & 3.7192e-2 & 1.4807e-1 \\
  2 & 9.0848e-2 & 2.6154e-2 & -1.6323e-2 & 6.5384e-3 & 2.2712e-2 & 8.5073e-2 \\
  3 & 5.8932e-2 & 1.7123e-2 & -1.0156e-2 & 4.2809e-3 & 1.4733e-2 & 5.3653e-2 
\end{tabular} 
\end{ruledtabular} 
\end{table} 

The first term on the right of Eq.~(\ref{gamma_intro}) is the $r$-mode contribution to the total accumulated phase shift, the one that was previously calculated by Flanagan and Racine \cite{flanagan-racine:07}. This term is proportional to $\sin^2\iota(\cos\iota+1)^2$, which is maximized when $\iota = \pi/3$. As Table~\ref{tab:uvalues_intro} indicates, the associated number $\gothu^2_\bullet$ is negative, which implies that the $r$-mode makes a negative contribution to the phase shift. With the sign convention used here\footnote{The sign convention is the one of Flanagan and Racine \cite{flanagan-racine:07}, for whom a positive $\Delta \Phi_{\rm GW}$ corresponds to a shorter inspiral. This convention is opposite to the one adopted by Lai in Ref.~\cite{lai:94}.}, this means that the resonant tidal interaction provides energy to the orbital motion, thereby increasing the orbital radius and prolonging the inspiral. This behavior has to do with the well-known fact that the $r$-mode is subjected to the Chandrasekhar-Friedman-Schutz instability \cite{chandrasekhar:70, friedman-schutz:78b, andersson:98, friedman-morsink:98}; the increase in orbital energy comes at the expense of the star's rotational energy.

The other terms on the right of Eq.~(\ref{gamma_intro}) are the contributions from the remaining inertial modes. They are all positive, which means that the resonant tidal interaction removes energy from the orbital motion, thereby decreasing the orbital radius and shortening the inspiral. The contribution from the $(m=1, n = \mbox{I})$ mode is proportional to $(\cos\iota + 1)^2(2\cos\iota - 1)^2$, which is maximized when $\iota = 0$, that is, when the angular-momentum vectors are aligned. The contribution from the $(m=1, n = \mbox{II})$ mode is proportional to $(\cos\iota - 1)^2(2\cos\iota + 1)^2$; this is maximized when $\iota = \pi$, for anti-aligned angular-momentum vectors. Finally, the contribution from the $(m=0, n = \mbox{I})$ pair of modes is proportional to $\sin^2\iota \cos^2\iota$, which is maximized when $\iota = \pi/4$ or $\iota = 3\pi/4$. 

\begin{figure} 
\includegraphics[width=0.7\linewidth]{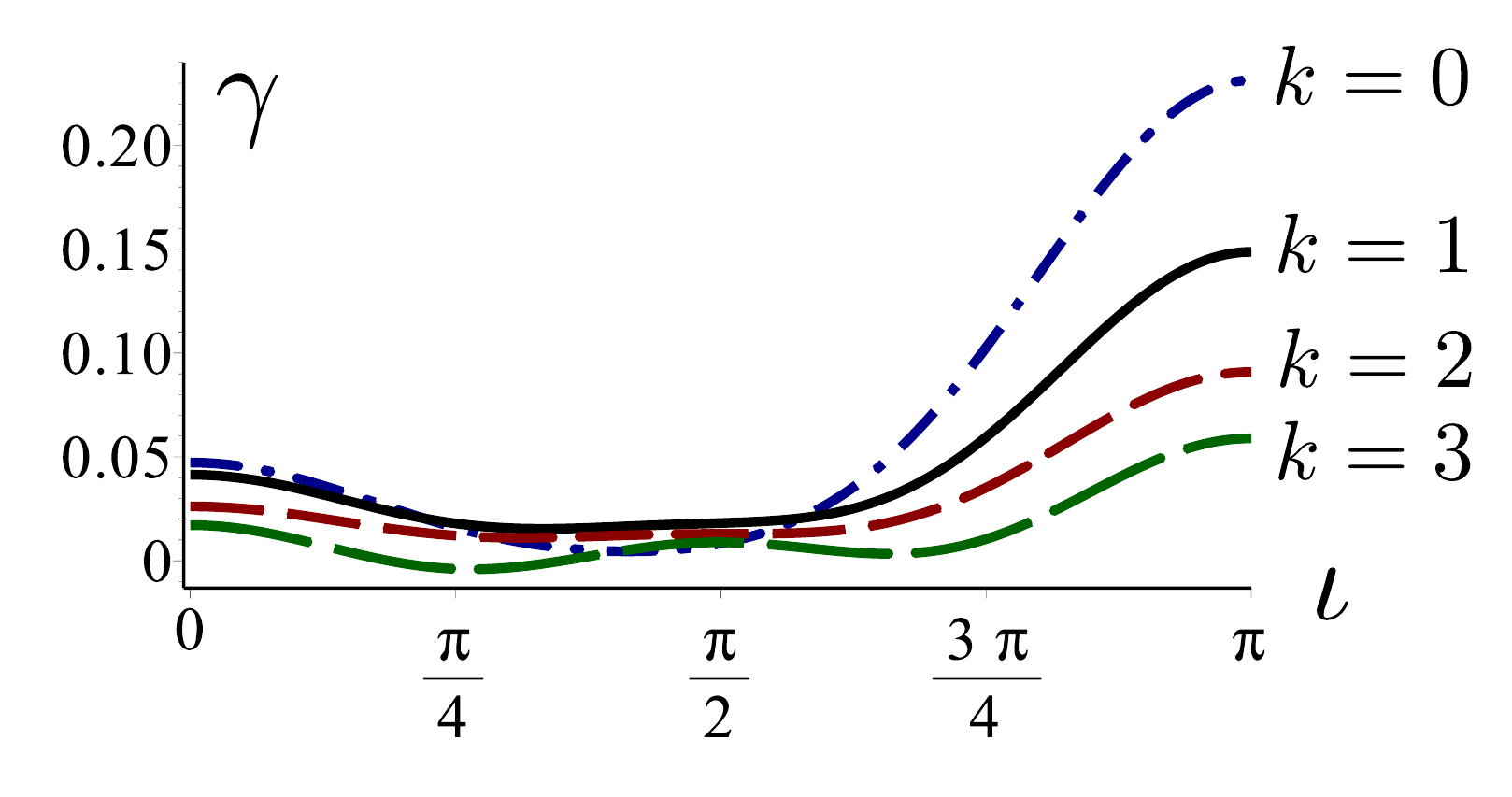}
\caption{Gravitational-wave phase shift $\gamma$ as a function of inclination angle $\iota$, for selected values of the density model parameter $k$. As stated in the text, the density model with $k=1$ gives a crude but effective description of a neutron star with a realistic equation of state.}   
\label{fig:phaseshift_intro} 
\end{figure}

Plots of $\gamma(\iota)$, the sum of all four contributions to the gravitational-wave phase shift, are displayed in Fig.~\ref{fig:phaseshift_intro} for selected values of $k$. The main observation is that the phase shift is maximized when $\iota = \pi$, that is, when the spin is anti-aligned with orbital angular momentum. In this situation the only contributing mode is the one labelled by $(m = 1, n = {\rm II})$ --- the inertial mode with a negative frequency --- and the phase shift becomes $\gamma(\iota = \pi) = 4 \gothu^1_{\rm II}$. On the other hand, when the spin is aligned with the orbital angular momentum, so that $\iota = 0$, the only contributing mode is $(m = 1, n = {\rm I}$), and we have that $\gamma(\iota = 0) = 4 \gothu^1_{\rm I}$. The values of $\gamma$ at $\iota = \pi$ and $\iota = 0$ are listed in Table~\ref{tab:uvalues_intro}. 

It can also be observed from Table~\ref{tab:uvalues_intro} that $\gothu_n^m$ decreases (in absolute value) with increasing $k$: stellar models that are more centrally dense produce a smaller phase shift. This behavior is to be expected, on the grounds that everything else being equal, a star that is more centrally dense develops a smaller current quadrupole moment $J^{jk}$, and therefore undergoes a weaker tidal interaction. 

With realistic values for $M$, $M'$, $\Omega$, and $R$, and for density models that are neither too uniform nor centrally dense, $\Delta \Phi_{\rm GW}$ can be expected to be in an interval between $0.02$ and $0.04$ radians when the spin is approximately aligned with the orbital angular momentum, or in an interval between $0.08$ and $0.16$ radians when the spin is approximately anti-aligned. If the companion also is a rotating neutron star, then both stars will participate in the resonant tidal interaction, and the total accumulated phase shift will be multiplied by two (assuming that the spins are comparable). The chosen fiducial value for $R$ --- $10\ \mbox{km}$ --- corresponds to a fairly soft equation of state that produces a relatively small star; a stiffer equation of state would return a larger stellar radius. A more optimistic choice of $13\ \mbox{km}$ for the fiducial value leads to an enhancement of the phase shift by a factor of $2.9$ --- thanks to the scaling with $R^4$, a little optimism goes a long way. On the other hand, the selection of $100\ \mbox{Hz}$ as a fiducial value for $\Omega/(2\pi)$ might be overly optimistic, given that the fastest known pulsar rotates at a rate of about $44\  \mbox{Hz}$\cite{burgay-etal:03}. A reduction of the fiducial value to $40\ \mbox{Hz}$ would reduce the phase shift by a factor of $0.54$ --- because of the scaling with $\Omega^{2/3}$, strong pessimism comes with a relatively small impact.   

With good constraints on the masses and spins (and therefore on the inclination angle $\iota$) gathered from the inspiral waveform, $\Delta \Phi_{\rm GW}$ provides information regarding the star's internal structure through the quantity $\gamma(\iota) R^4$, which depends on the equation of state. Just how well this information can be extracted from gravitational-wave measurements, and precisely what can be inferred about the equation of state, are important questions that lie beyond the scope of the paper. For the time being I shall limit myself to a statement of optimism regarding future measurements of the gravitomagnetic tidal resonances: 

In their recent study of the excitation of interface modes during binary inspirals, Pan {\it et al.}\ \cite{pan-etal:20} estimate that for an event comparable to GW170817 \cite{GW170817:17}, the gravitational-wave phase shift $\Delta \Phi_{\rm GW}$ can be measured by the current LIGO/Virgo facility within an uncertainty comparable to $0.1$ radians. This, unfortunately, is about the size of the effect in the most favorable circumstances. But these authors also estimate that for a similar event, measured with a signal-to-noise ratio of 100 by LIGO A+ \cite{LIGOAplus}, the uncertainty in the phase shift will decrease by an order of magnitude, a sufficient improvement to reveal the phenomenon. A similar conclusion was reached by Yu {\it et al.}\ \cite{yu-etal:18} in their proposal for a detector upgrade (LIGO-LF) that dramatically improves the sensitivity at low frequencies. I take this as a strong indication that the gravitational-wave phase shift accumulated during a gravitomagnetic tidal resonance will be within reach in the current decade, at the cost of an incremental improvement of the LIGO/Virgo instruments. Thinking ahead, the resonant tidal interaction will have to be incorporated in templates when the waves are measured by the next generation of detectors, such as Cosmic Explorer \cite{CosmicExplorer, reitze-etal:19} and the Einstein Telescope \cite{EinsteinTelescope, punturo:10}. 

\subsection{Some fine print: approximations made in this paper}

The results summarized in the preceding subsection were obtained by formulating various approximations to the correct and complete physics of the problem. I state and discuss these assumptions here. 

First, the gravitomagnetic tidal field acting on the neutron star is described to leading order only in a post-Newtonian expansion in powers of $v'/c$, where $v'$ is the orbital velocity and $c$ is the speed of light. At this leading, first post-Newtonian order, it derives from a vector potential $U_j$ created by the mass current associated with the companion's orbital motion. The vector potential is essentially the time-space part of the metric tensor. 

Second, the fluid dynamics is described within a purely Newtonian framework. While the external forces have a post-Newtonian origin, and the resulting perturbation of the fluid configuration is also of the first post-Newtonian order, the perturbation is nevertheless calculated with the Newtonian fluid equations. There is no inconsistency with this approach, because the focus is on gravitomagnetic tidal effects; other post-Newtonian corrections are ignored. 

Third, the unperturbed ``neutron star'' is actually a Newtonian configuration of perfect fluid, self-gravitating and rotating rigidly with an angular velocity $\Omega$, governed by a barotropic equation of state of the form $p = p(\rho)$, where $p$ is the pressure and $\rho$ the mass density. The perturbed star is assumed to obey the same equation of state. Deviations from the barotropic form, associated with composition gradients and/or a nonzero temperature, would lead to buoyancy effects in the inertial modes, and these are neglected. Similarly, no attempt is made to account for the influence of the superfluid core on the modes (for a detailed study on $r$-modes, see Ref.~\cite{lindblom-mendell:00}). 

Fourth, it is assumed that the star is rotating slowly, in the sense that $\Omega^2 \ll GM/R^3$, where $G$ is the gravitational constant. All centrifugal corrections to the stellar structure are neglected in this approximation; the surfaces of constant density and pressure are spheres. The inertial modes and their frequencies are calculated within this approximation. 

Fifth, the binary's inspiral is described to leading order only in a post-Newtonian expansion of the radiation reaction force. At this order, the orbital evolution is a second-and-a-half post-Newtonian effect, associated with the loss of orbital energy and angular momentum to gravitational waves. The orbit is taken to be circular.

Sixth, the treatment provided here of the development of gravitomagnetic tidal resonances is based on the assumption that the transit time through a resonance is short compared with the radiation-reaction time scale. The ratio of time scales is denoted $\epsilon$, and many expressions below neglect fractional corrections of order $\epsilon$. Because $\epsilon \sim 10^{-2}$ for binaries of interest [see Eq.~(\ref{epsilon_def}) below], this provides a perfectly adequate approximation. A consequence of this hierarchy of time scales is that the four resonances are well separated in time, since they each have a distinct frequency.  

Seventh, dissipation of the velocity perturbation through the coupling of an inertial mode to the star's shear and bulk viscosity, and through the mode's own emission of gravitational waves, is neglected. In this treatment, the only source of dissipation in the system is the production of gravitational waves by the binary's orbital motion.   

All these approximations imply that the numbers listed in Table~\ref{tab:uvalues_intro} can be expected to be accurate only up to about 10 or 20 percent. The dominant source of error is the Newtonian treatment of the neutron-star interior and the Newtonian calculation of inertial modes. Promotion to general relativity will have to await future work.  

\subsection{Organization of this paper}

In the first part of the paper --- Secs.~\ref{sec:governing} to \ref{sec:solution} --- I calculate the response of a slowly rotating star to an applied gravitomagnetic tidal field, over a time interval that is long compared with the orbital period, but short compared with the transit time through a resonance (and therefore also short compared with the radiation-reaction time scale). On this short-term view the binary's orbital radius can be taken to be fixed, and the frequency of the tidal field is approximately constant. The equations that govern the dynamics of a barotropic perfect fluid are reviewed in Sec.~\ref{sec:governing}, and the gravitomagnetic tidal force is introduced in Sec.~\ref{sec:gravitomagnetic}. The explicit form of the perturbation equations is obtained in Sec.~\ref{sec:perturbation}, and solutions to these equations are constructed in Sec.~\ref{sec:solution}. The main goal in this first part of the paper is to establish that the star's tidal response can be described in terms of dimensionless current quadrupole moments $\gothj_m(\omega)$, one for each value of $m = \{0, 1, 2\}$, which are functions of the orbital frequency $\omega$. These quantities are defined in Eq.~(\ref{j_components}) below, they are plotted in Figs.~\ref{fig:j_m2}, \ref{fig:j_m1}, and \ref{fig:j_m0}, and the current quadrupole moment tensor $J^{jk}$ is related to them via Eqs.~(\ref{J_decomposed}) and (\ref{J_components}).  

The second part of the paper --- Sec.~\ref{sec:oscillator} --- is also concerned with a short-term view of the gravitomagnetic tidal interaction. It is devoted to a representation of the velocity perturbation in terms of a sum over normal modes. While an exact expression would require a sum over an infinite number of modes, the main purpose of Sec.~\ref{sec:oscillator} is to establish that an excellent approximation results from the inclusion of just four modes, namely the inertial modes introduced previously. The four-mode approximation is at once suggested and validated by Figs.~\ref{fig:j_m2}, \ref{fig:j_m1}, and \ref{fig:j_m0}. The suggestion comes from the fact that the reduced current quadrupole moments $\gothj_m(\omega)$ feature simple poles at eigenfrequencies of the inertial modes; the validation comes from the agreement between the solid curves (the outcome of the four-mode approximation) and the open diamonds (the result of the direct calculation presented in Sec.~\ref{sec:solution}).    

In the third part of the paper --- Sec.~\ref{sec:dynamics} --- I turn to a long-term view in which the relevant time interval is of the order of the radiation-reaction time. In this view, the orbital radius decreases steadily in response to the radiation-reaction force, and at some point in the inspiral the orbital frequency momentarily becomes equal to a mode's  eigenfrequency. A succession of four resonances takes place, during which the orbital motion is significantly affected by its coupling with each mode; this coupling is mediated by the current quadrupole moment $J^{jk}$. I describe the inspiral, calculate how each mode grows during the development of a resonance, show how this affects the orbital motion, and obtain the gravitational-wave phase shift associated with each resonance. The final result was presented in  Eqs.~(\ref{DeltaPhi_intro}) and (\ref{gamma_intro}).    

The paper also contains an Appendix (Sec.~\ref{sec:phaseshift}), in which some (nonessential) technical matters are relegated.  

The developments in the main text rely heavily on the Lagrangian theory of fluid perturbations supplied by Friedman and Schutz \cite{friedman-schutz:78a}, as well as the extensions contributed by Schenk {\it et al.}\ \cite{schenk-etal:01}. I also made use of the methods introduced by Lockitch and Friedman \cite{lockitch-friedman:99} to calculate the inertial modes of slowly rotating, barotropic stars, and those devised by Flanagan and Racine \cite{flanagan-racine:07} to determine the dynamical impact of a gravitomagnetic tidal resonance.   

\section{Fluid equations} 
\label{sec:governing} 

We (you and I, working together) consider a star of mass $M$ and radius $R$, rotating slowly and rigidly with an angular velocity $\Omega$, perturbed by a gravitomagnetic tidal field described by a post-Newtonian vector potential $U_a$. The slow-rotation assumption implies that the unperturbed star is approximately spherically symmetric, with all centrifugal deformations neglected. We take the star to be described by a perfect fluid with a barotropic equation of state $p = p(\rho)$, where $p$ is the pressure and $\rho$ the mass density. The physics of the unperturbed stellar interior is described within the framework of Newtonian fluid mechanics and gravitation. The first governing equation is Euler's equation    
\begin{equation} 
0 = E_a := \partial_t v_a + v^b \nabla_b v_a - \nabla_a (U - h), 
\label{euler} 
\end{equation} 
where $v^a$ is the velocity field, $U$ the Newtonian gravitational potential, and $h$ the specific enthalpy, defined by $dh = \rho^{-1}\, dp$. Euler's equation is written in covariant form, and can therefore be formulated in any coordinate system; the same remark applies to all equations in this section. For the unperturbed state we have that $v^a = \Omega\, \phi^a$, where $\phi^a$ is the azimuthal Killing vector, the time-derivative term vanishes in Eq.~(\ref{euler}), and the term quadratic in $\Omega$ is neglected within the slow-rotation approximation; the equation of hydrostatic equilibrium reduces to $\nabla_a (U - h) = 0$. The second governing equation is the continuity equation 
\begin{equation} 
0 = C := \partial_t \rho + \nabla_a (\rho v^a),  
\label{continuity} 
\end{equation} 
which is trivially satisfied for a static and axisymmetric distribution of mass. A third governing equation is Poisson's equation, 
\begin{equation} 
0 = P := \nabla^2 U +  4\pi G \rho,   
\label{poisson} 
\end{equation} 
which determines the gravitational potential.  

The perturbation takes its origin in post-Newtonian gravity. We suppose that the star is placed within a time-dependent gravitomagnetic tidal field described by a vector potential $U_a$, and that the fluid is therefore subjected to a perturbing force density given by $\rho f_a$, where  
\begin{equation} 
f_a = \frac{4}{c^2} \bigl[ \partial_t U_a + v^b (\nabla_b U_a - \nabla_a U_b) \bigr]. 
\label{force_density} 
\end{equation} 
This is Eq.~(8.119) of Ref.~\cite{poisson-will:14}, after discarding all terms that are irrelevant for our purposes, because they do not involve the vector potential.  

The tidal force creates a perturbation in the fluid, described by the Eulerian changes $\{ \delta v^a, \delta\rho, \delta p, \delta h, \delta U \}$. The perturbation of Euler's equation is 
\begin{equation} 
\delta E_a = f_a, 
\label{dEuler} 
\end{equation} 
where 
\begin{equation} 
\delta E_a := \partial_t \delta v_a + \delta v^b\, \nabla_b v_a + v^b \nabla_b \delta v_a 
- \nabla_a (\delta U - \delta h). 
\label{dEa} 
\end{equation} 
The velocity perturbation can be related to a Lagrangian displacement vector $\xi^a$ by 
\begin{equation} 
\delta v^a = \partial_t \xi^a - \Lie_\xi v^a = (\partial_t + \Lie_v) \xi^a  
= (\partial_t + \Omega\, \partial_\phi) \xi^a,  
\label{xi_vs_dv} 
\end{equation}  
where $\Lie$ is the Lie-derivative operator. The perturbed continuity equation implies that 
\begin{equation} 
\delta \rho = -\nabla_a (\rho \xi^a), 
\label{drho} 
\end{equation} 
and $\delta p$, $\delta h$ can be obtained from $\delta \rho$ and the equation of state. The perturbed Poisson equation is 
\begin{equation} 
\nabla^2 \delta U + 4\pi G \delta\rho = 0, 
\label{dPoisson} 
\end{equation} 
and it determines $\delta U$. 

The perturbation equations come with the requirement that $\Delta h = 0$ at the stellar surface, where $\Delta h = \delta h + \xi^a \nabla_a h$ is the Lagrangian change in the specific enthalpy. With $dh/dr = -GM/R^2$ at $r=R$, the boundary condition becomes 
\begin{equation} 
\delta h(r = R) = \frac{GM}{R^2} r_a \xi^a (r=R), 
\label{surface_condition} 
\end{equation} 
where $r_a := \nabla_a r$ is the radial unit vector. 

\section{Gravitomagnetic tidal force} 
\label{sec:gravitomagnetic} 

\subsection{Tidal quadrupole moment} 

The gravitomagnetic tidal field is characterized by a symmetric-tracefree quadrupole-moment tensor ${\cal B}_{jk}(t)$, which gives rise to the vector potential 
\begin{equation} 
U_j = -\frac{1}{6} \epsilon_{jkp} {\cal B}^p_{\ q} x^k x^q, 
\label{U_vs_B} 
\end{equation} 
where $\epsilon_{jkp}$ is the permutation symbol, and $x^j := (x,y,z)$ are Cartesian coordinates attached to the star's center-of-mass; the $z$-direction is aligned with the rotation axis.\footnote{While all equations in Sec.~\ref{sec:governing} were written in covariant form, the expression for the vector potential is restricted to Cartesian coordinates. To distinguish covariant equations from equations valid only in Cartesian coordinates, we use indices $a, b, c, \ldots$ in covariant equations, and indices $j, k, p, \ldots$ in Cartesian equations.} Equation (\ref{U_vs_B}) follows from the requirements that the vector potential must satisfy Laplace's equation $\nabla^2 U_j = 0$ and the gauge condition $\partial_j U^j = 0$. The potential is substituted within Eq.~(\ref{force_density}) to obtain the tidal force density. 

The tidal field is created by the orbital motion of a companion body. This companion has a mass $M'$, and the orbital radius is denoted $p$ (the standard symbol for semi-latus rectum). The orbital angular velocity is
\begin{equation} 
\omega = \sqrt{\frac{GM_{\rm tot}}{p^3}}, 
\label{orbital} 
\end{equation} 
where $M_{\rm tot} := M + M'$ is the binary's total mass. The orbital velocity is $v' = p\, \omega$. 

In a generic orientation, the normal to the orbital plane is directed along the vector $\bm{l}$, the direction of the companion (with respect to the origin of the coordinate system) is $\bm{n}$, and the direction of the orbital velocity vector is $\bm{\lambda}$. These vectors have components $\bm{l} = [0,-\sin\iota,\cos\iota]$, $\bm{n} = [\cos\omega t, \cos\iota\, \sin\omega t, \sin\iota\, \sin\omega t]$, and $\bm{\lambda} = [-\sin\omega t, \cos\iota\, \cos\omega t, \sin\iota\, \sin\omega t]$, where $\iota$ is the inclination angle between the normal to the orbital plane and the star's rotation axis. The vectors have a unit length, they are mutually orthogonal, and $\bm{l} = \bm{n} \times \bm{\lambda}$.

The tidal quadrupole moment is given by \cite{taylor-poisson:08} 
\begin{equation} 
{\cal B}_{jk} = 3 {\cal B}_0 \bigl( l_j n_k + n_j l_k \bigr), \qquad 
{\cal B}_0 := \frac{GM'v'}{p^3} = \frac{GM'}{(GM_{\rm tot})^{2/3}}\, \omega^{7/3}.   
\label{Bjk} 
\end{equation} 
An explicit listing of components is 
\begin{subequations} 
\label{B_listing} 
\begin{align} 
{\cal B}_{xx} &= 0, \\ 
{\cal B}_{xy} &= -3 {\cal B}_0 \sin\iota\, \cos(\omega t), \\
{\cal B}_{xz} &= 3 {\cal B}_0 \cos\iota\, \cos(\omega t), \\
{\cal B}_{yy} &= -6 {\cal B}_0 \sin\iota\cos\iota\, \sin(\omega t), \\
{\cal B}_{yz} &= 3 {\cal B}_0 (2\cos^2\iota-1)\, \sin(\omega t), \\
{\cal B}_{zz} &= 6 {\cal B}_0 \sin\iota\cos\iota\, \sin(\omega t). 
\end{align} 
\end{subequations} 

\subsection{Force density} 

We substitute Eqs.~(\ref{B_listing}) within Eq.~(\ref{U_vs_B}), and this within the force density of Eq.~(\ref{force_density}). For our purposes below it is useful to express the result in spherical coordinates, as
\begin{equation}
f_a = \sum_{m=-2}^2 f_a^m e^{im\phi} e^{-i\omega t} + \mbox{cc}, 
\label{fa_decomposed} 
\end{equation}
where ``cc'' denotes the complex conjugate of the preceding expression. The coefficients are given by 
\begin{subequations}
\label{fa_listing_m2} 
\begin{align} 
f^{\pm 2}_r &= \frac{3i}{2} \frac{{\cal B}_0}{c^2} \sin\iota (\cos\iota \pm 1) \Omega \sin^2\theta\cos\theta\, r^2, \\
f^{\pm 2}_\theta &= \frac{i}{2} \frac{{\cal B}_0}{c^2} \sin\iota (\cos\iota \pm 1) (-3\Omega \sin^2\theta
\pm \omega) \sin\theta\, r^3, \\
f^{\pm 2}_\phi &= -\frac{1}{2} \frac{{\cal B}_0}{c^2} \sin\iota (\cos\iota \pm 1) \omega \sin^2\theta\cos\theta\, r^3, 
\end{align}
\end{subequations} 
\begin{subequations}
\label{fa_listing_m1} 
\begin{align} 
f^{\pm 1}_r &= \pm \frac{3}{2} \frac{{\cal B}_0}{c^2} (\cos\iota \pm 1)(2\cos\iota \mp 1) \Omega (2\cos^2\theta - 1)\sin\theta\, r^2, \\
f^{\pm 1}_\theta &= \pm \frac{1}{2} \frac{{\cal B}_0}{c^2} (\cos\iota \pm 1)(2\cos\iota \mp 1) (-6\Omega\sin^2\theta \pm \omega ) \cos\theta\, r^3, \\
f^{\pm 1}_\phi &= \pm \frac{i}{2} \frac{{\cal B}_0}{c^2} (\cos\iota \pm 1)(2\cos\iota \mp 1) \omega \sin\theta
(2\cos^2\theta - 1) r^3,
\end{align}
\end{subequations} 
and
\begin{subequations}
\label{fa_listing_m0} 
\begin{align} 
f^0_r &= -9i\frac{{\cal B}_0}{c^2} \sin\iota \cos\iota\, \Omega \sin^2\theta \cos\theta\, r^2, \\
f^0_\theta &= -3i\frac{{\cal B}_0}{c^2} \sin\iota \cos\iota\, \Omega \sin\theta (3\cos^2\theta-1)\, r^3, \\
f^0_\phi &= 3\frac{{\cal B}_0}{c^2} \sin\iota \cos\iota\, \omega \sin^2\theta \cos\theta\, r^3. 
\end{align}
\end{subequations} 

\subsection{Curl of the force density} 

In subsequent developments we shall also need the curl of the force density, 
\begin{equation} 
q^a := \epsilon^{abc} \nabla_b f_c,  
\label{q_def} 
\end{equation}    
where $\epsilon_{abc}$ is the completely antisymmetric Levi-Civita tensor; in spherical coordinates we have that $\epsilon_{r\theta\phi} = r^2\sin\theta$. We decompose $q^a$ as
\begin{equation}
q^a = \sum_{m=-2}^2 q^a_m e^{im\phi} e^{-i\omega t} + \mbox{cc},
\label{qa_decomposed} 
\end{equation}
with
\begin{subequations}
\label{qa_listing_m2}
\begin{align}
q^r_{\pm 2} &= \mp \frac{3}{2} \frac{{\cal B}_0}{c^2} \sin\iota (\cos\iota \pm 1)(2\Omega \mp \omega) \sin^2\theta\, r, \\
q^\theta_{\pm 2} &= \mp \frac{3}{2} \frac{{\cal B}_0}{c^2} \sin\iota (\cos\iota \pm 1)(2\Omega \mp \omega) \sin\theta\cos\theta, \\
q^\phi_{\pm 2} &= -\frac{3i}{2} \frac{{\cal B}_0}{c^2} \sin\iota (\cos\iota \pm 1)(2\Omega \mp \omega),
\end{align}
\end{subequations}
\begin{subequations}
\label{qa_listing_m1}
\begin{align}
q^r_{\pm 1} &= 3i \frac{{\cal B}_0}{c^2}  (\cos\iota \pm 1)(2\cos\iota \mp 1) (\Omega \mp \omega) \sin\theta\cos\theta\, r, \\
q^\theta_{\pm 1} &= \frac{3i}{2} \frac{{\cal B}_0}{c^2}  (\cos\iota \pm 1)(2\cos\iota \mp 1) (\Omega \mp \omega) (2\cos^2\theta-1), \\
q^\phi_{\pm 1} &= \mp \frac{3}{2} \frac{{\cal B}_0}{c^2}  (\cos\iota \pm 1)(2\cos\iota \mp 1) (\Omega \mp \omega) \cot\theta, 
\end{align}
\end{subequations}
and 
\begin{subequations}
\label{qa_listing_m0}
\begin{align}
q^r_0 &= 3 \frac{{\cal B}_0}{c^2} \sin\iota \cos\iota\, \omega (3\cos^2\theta - 1) r, \\
q^\theta_0 &= -9 \frac{{\cal B}_0}{c^2} \sin\iota \cos\iota\, \omega \sin\theta\cos\theta, \\
q^\phi_0 &= 0.
\end{align}
\end{subequations}

\subsection{Spherical-harmonic decompositions}
\label{subsec:harmonics}

Subsequent developments will require a set of vectorial harmonics to decompose vector fields such as $\delta v^a$ and $f^a$. The first member of the set consists of the radial harmonics
\begin{equation}
r^a Y_\ell^m,
\end{equation}
the union of the usual scalar harmonics $Y_\ell^m$ with the unit radial vector $r^a := \nabla^a r$. In Cartesian coordinates, this vector is given by $r^j = x^j/r$; in spherical coordinates, its only nonvanishing component is $r^r = 1$. The second member of the set consists of the polar harmonics
\begin{equation}
(Y_\ell^m)^a := \nabla^a Y_\ell^m.
\end{equation}
And the third member of the set consists of the axial harmonics
\begin{equation}
(X_\ell^m)^a := \epsilon^{abc} (\nabla_b Y_\ell^m) r_c, 
\end{equation}
the cross product between the polar harmonics and the unit radial vector. 

The scalar harmonics are normalized according to 
\begin{equation}
\int \bar{Y}_\ell^m Y_{\ell'}^{m'}\, d\Omega = \delta_{\ell\ell'} \delta_{mm'}, 
\label{ortho1} 
\end{equation} 
where an overbar indicates complex conjugation, and $d\Omega := \sin\theta\, d\theta d\phi$ is the element of solid angle. It follows from this and the definitions of the polar and axial harmonics that 
\begin{subequations} 
\label{ortho2} 
\begin{align} 
\int (\bar{Y}_\ell^m)_a (Y_{\ell'}^{m'})^a\, d\Omega &= \frac{\ell(\ell+1)}{r^2} \delta_{\ell\ell'} \delta_{mm'}, \\ 
\int (\bar{X}_\ell^m)_a (X_{\ell'}^{m'})^a\, d\Omega &= \frac{\ell(\ell+1)}{r^2} \delta_{\ell\ell'} \delta_{mm'}.  
\end{align} 
\end{subequations} 
We also have that the polar harmonics are orthogonal to the axial harmonics, and both are orthogonal to the radial harmonics. We adhere to the convention $Y_\ell^{-m} = (-1)^m \bar{Y}_\ell^m$. 

The decomposition of $f_a$ in vector harmonics is accomplished by
\begin{equation}
f_a = \sum_{\ell m} \biggl[ \frac{1}{r} (f_\ell^m)^{\rm R}\, r_a Y_\ell^m
+ (f_\ell^m)^{\rm P} (Y_\ell^m)_a  + (f_\ell^m)^{\rm A} (X_\ell^m)_a \biggr] e^{-i\omega t} + \mbox{cc},
\label{fa_vharm}
\end{equation}
where the nonvanishing coefficients are
\begin{subequations} 
\label{fa_radial} 
\begin{align} 
(f_3^{\rm \pm 2})^{\rm R} &= \frac{2i}{35} \sqrt{210\pi} \frac{{\cal B}_0}{c^2} \sin\iota(\cos\iota \pm 1) \Omega\, r^3, \\
(f_3^{\rm \pm 1})^{\rm R} &= -\frac{8}{35} \sqrt{21\pi} \frac{{\cal B}_0}{c^2} (\cos\iota \pm 1)(2\cos\iota \mp 1) \Omega\, r^3, \\
(f_1^{\pm 1})^{\rm R} &= \frac{3}{5} \sqrt{6\pi} \frac{{\cal B}_0}{c^2} (\cos\iota \pm 1)(2\cos\iota \mp 1) \Omega\, r^3, \\
(f_3^0)^{\rm R} &= \frac{36i}{35} \sqrt{7\pi} \frac{{\cal B}_0}{c^2} \sin\iota\cos\iota\, \Omega\, r^3, \\
(f_1^0)^{\rm R} &= -\frac{12i}{5} \sqrt{3\pi} \frac{{\cal B}_0}{c^2} \sin\iota\cos\iota\, \Omega\, r^3, 
\end{align}
\end{subequations}
\begin{subequations} 
\label{fa_polar} 
\begin{align} 
(f_3^{\rm \pm 2})^{\rm P} &= \frac{2i}{105} \sqrt{210\pi} \frac{{\cal B}_0}{c^2} \sin\iota(\cos\iota \pm 1) \Omega\, r^3, \\
(f_3^{\rm \pm 1})^{\rm P} &= -\frac{8}{105} \sqrt{21\pi} \frac{{\cal B}_0}{c^2} (\cos\iota \pm 1)(2\cos\iota \mp 1) \Omega\, r^3, \\
(f_1^{\pm 1})^{\rm P} &= \frac{1}{5} \sqrt{6\pi} \frac{{\cal B}_0}{c^2} (\cos\iota \pm 1)(2\cos\iota \mp 1) \Omega\, r^3, \\
(f_3^0)^{\rm P} &= \frac{12i}{35} \sqrt{7\pi} \frac{{\cal B}_0}{c^2} \sin\iota\cos\iota\, \Omega\, r^3, \\
(f_1^0)^{\rm P} &= -\frac{4i}{5} \sqrt{3\pi} \frac{{\cal B}_0}{c^2} \sin\iota\cos\iota\, \Omega\, r^3,
\end{align}
\end{subequations}
and
\begin{subequations} 
\label{fa_axial} 
\begin{align} 
(f_2^{\pm 2})^{\rm A} &= \mp \frac{1}{15} \sqrt{30\pi} \frac{{\cal B}_0}{c^2} \sin\iota (\cos\iota \pm 1) (2\Omega \mp \omega)\, r^3, \\
(f_2^{\pm 1})^{\rm A} &= \mp \frac{i}{15} \sqrt{30\pi} \frac{{\cal B}_0}{c^2} (\cos\iota \pm 1)(2\cos\iota \mp 1) (\Omega \mp \omega)\, r^3, \\
(f_2^0)^{\rm A} &= \frac{2}{5} \sqrt{5\pi} \frac{{\cal B}_0}{c^2} \sin\iota\cos\iota\, \omega\, r^3.  
\end{align}
\end{subequations}
The superscripts on $(f_\ell^m)$ stand for ``Radial'', ``Polar'', and ``Axial''.

An observation that will be exploited in the sequel is that 
\begin{equation}
(f_\ell^m)^{\rm R} = 3 k_\ell^m\, r^3, \qquad
(f_\ell^m)^{\rm P} = k_\ell^m\, r^3 
\label{RP-proportionality}
\end{equation}
for all relevant values of $\ell$ and $m$, with the same constant $k_\ell^m$ in both expressions. 

We shall not require a decomposition of $q^a$ in vectorial harmonics. 
 
\section{Perturbation equations} 
\label{sec:perturbation} 

We now examine the system of perturbation equations for the variables $\{ \delta v^a, \delta \rho, \delta p, \delta h, \delta U\}$. This consists of Euler's equation (\ref{dEuler}), the continuity equation (\ref{drho}), Poisson's equation (\ref{dPoisson}), and the surface condition of Eq.~(\ref{surface_condition}). Following the general strategy outlined by Lockitch and Friedman \cite{lockitch-friedman:99}, we shall show that the perturbation is dominated by $\delta v^a$. We continue to assume that $\Omega$ is small, so that all centrifugal effects can be neglected, and we further assume that $\omega$ and $\Omega$ are of the same order of magnitude.   

\subsection{Velocity perturbation} 

Following Lockitch and Friedman \cite{lockitch-friedman:99}, we eliminate $\delta U$ and $\delta h$ from Euler's equation (\ref{dEuler}) by taking the curl of both sides. With the definition of Eq.~(\ref{q_def}), this gives
\begin{equation}
Q^a = q^a,
\label{curlEuler} 
\end{equation} 
where $Q^a := \epsilon^{abc} \nabla_b \delta E_c$. Equation (\ref{curlEuler}) provides only two independent equations, because $\nabla_a Q^a = 0 = \nabla_a q^a$ (the divergence of a curl is always zero). 

The explicit form of Eq.~(\ref{curlEuler}) and the decomposition of Eq.~(\ref{qa_decomposed}) imply that $\delta v^a$ admits the decomposition
\begin{equation}
\delta v^a = \sum_{m=-2}^2 \delta v^a_m e^{-i\omega t} + \mbox{cc},
\label{va_decomposed1} 
\end{equation}
where each $\delta v^a_m$ can be further decomposed in vectorial harmonics. We incorporate the scaling and numerical factors implied by Eqs.~(\ref{qa_listing_m2}), (\ref{qa_listing_m1}), and (\ref{qa_listing_m0}), and write
\begin{subequations}
\label{va_decomposed2}
\begin{align}
\delta v^a_{\pm 2} &= \frac{{\cal B}_0}{c^2} R^3 \sin\iota(\cos\iota \pm 1)(2 \mp \omega/\Omega) \Biggl[
\mp \sum_{\ell=3}^{\rm odd} \frac{1}{r} A_\ell^{\pm 2}\, r^a Y_\ell^{\pm 2}
\mp \sum_{\ell=3}^{\rm odd} B_\ell^{\pm 2}\, (Y_\ell^{\pm 2})^a
- i \sum_{\ell=2}^{\rm even} C_\ell^{\pm 2}\, (X_\ell^{\pm 2})^a \Biggr], \\
\delta v^a_{\pm 1} &= \frac{{\cal B}_0}{c^2} R^3 (\cos\iota \pm 1)(2\cos\iota \mp 1)(1 \mp \omega/\Omega) \Biggl[
i \sum_{\ell=1}^{\rm odd} \frac{1}{r} A_\ell^{\pm 1}\, r^a Y_\ell^{\pm 1}
+ i\sum_{\ell=1}^{\rm odd} B_\ell^{\pm 1}\, (Y_\ell^{\pm 1})^a
\mp \sum_{\ell=2}^{\rm even} C_\ell^{\pm 1}\, (X_\ell^{\pm 1})^a \Biggr], \\
\delta v^a_{0} &= \frac{{\cal B}_0}{c^2} R^3 \sin\iota\cos\iota\, (\omega/\Omega) \Biggl[
\sum_{\ell=1}^{\rm odd} \frac{1}{r} A_\ell^{0}\, r^a Y_\ell^{0}
+ \sum_{\ell=1}^{\rm odd} B_\ell^{0}\, (Y_\ell^{0})^a
- i \sum_{\ell=2}^{\rm even} C_\ell^{0}\, (X_\ell^{0})^a \Biggr],
\end{align}
\end{subequations}
where the radial functions $A_\ell^m(r)$, $B_\ell^m(r)$, and $C_\ell^m(r)$ are all dimensionless. The restrictions on the sums over $\ell$ (even or odd terms only) will be justified below on the basis of the explicit form of Eq.~(\ref{curlEuler}). The various factors of $\pm$ and $i$ in front of these sums are inserted to simplify the perturbation equations.  

Equation (\ref{curlEuler}) implies that the radial functions satisfy the identities
\begin{equation}
A_\ell^{-2}(\omega) = A_\ell^2(-\omega), \qquad
B_\ell^{-2}(\omega) = B_\ell^2(-\omega), \qquad
C_\ell^{-2}(\omega) = C_\ell^2(-\omega), 
\label{ABCidentities1}
\end{equation}
\begin{equation}
A_\ell^{-1}(\omega) = -A_\ell^1(-\omega), \qquad
B_\ell^{-1}(\omega) = -B_\ell^1(-\omega), \qquad
C_\ell^{-1}(\omega) = -C_\ell^1(-\omega),  
\label{ABCidentities2}
\end{equation}
and 
\begin{equation}
A_\ell^0(\omega) = A_\ell^0(-\omega), \qquad
B_\ell^0(\omega) = B_\ell^0(-\omega), \qquad
C_\ell^0(\omega) = -C_\ell^0(-\omega). 
\label{ABCidentities3}
\end{equation}
It also implies that all radial functions are real. 

\subsection{Dominance of the velocity perturbation} 

When we insert within Eq.~(\ref{dEuler}) the scaling factors for $\delta v^a$ indicated in Eq.~(\ref{va_decomposed2}),  we find that $\delta h/|\delta v^a|$ and $\delta U/|\delta v^a|$ must both be of the order of $\omega R$ or $\Omega R$, making $\delta h$ and $\delta U$ (and therefore $\delta \rho$ and $\delta p$) of relative order $\Omega$ compared with $\delta v^a$. Stated in a more meaningful way, we have that 
\begin{equation} 
\frac{\delta h/h}{\delta v/v} \sim \frac{\delta U/U}{\delta v/v}
\sim \frac{\omega\Omega}{GM/R^3} \ 
\mbox{or} \ \frac{\Omega^2}{GM/R^3}, 
\end{equation} 
where $\delta v := |\delta v^a|$ and $v := |v^a| \sim \Omega R$. The small-$\Omega$ assumption ensures that these ratios are small. This relative scaling between $\delta v^a$ and the remaining perturbation variables is the same as for the Lockitch-Friedman inertial modes \cite{lockitch-friedman:99}, and this motivates an expectation that the gravitomagnetic tidal field acts as a driving force for these modes. This expectation will be verified in Sec.~\ref{sec:oscillator}.      

\subsection{Continuity equation and surface condition} 

As was stated previously, Eq.~(\ref{curlEuler}) provides only two independent equations for the three components of $\delta v^a$. A third equation is provided by the continuity equation (\ref{drho}). We express the Lagrangian displacement as
\begin{equation}
\xi^a = \sum_{m=-2}^2 \xi^a_m e^{-i\omega t} + \mbox{cc},
\label{xi_decomposed}
\end{equation}
and use Eq.~(\ref{xi_vs_dv}) to relate $\xi^a_m$ to $\delta v^a_m$; we obtain
\begin{equation}
\xi^a_m = i(\omega - m\Omega)^{-1}\, \delta v^a_m.
\end{equation}
Making the substitutions within Eq.~(\ref{drho}), taking $\omega$ to be of the same order of magnitude as $\Omega$, and recalling that $\delta \rho$ is formally of order $\Omega$, we find that the continuity equation reduces to  
\begin{equation} 
\nabla_a (\rho \delta v^a) = O(\Omega^2). 
\label{div_rhov} 
\end{equation} 
The third equation is therefore $\nabla_a (\rho \delta v^a) = 0$, and after substituting Eqs.~(\ref{va_decomposed1}) and (\ref{va_decomposed2}) we eventually obtain\footnote{To arrive at this we made use of the relations $\nabla_a r^a = 2r^{-1}$, $\nabla_a (Y_\ell^m)^a = -\ell(\ell+1) r^{-2}\, Y_\ell^m$, and $\nabla_a (X_\ell^m)^a = 0$.}
\begin{equation} 
r \frac{dA_\ell^m}{dr} + \biggl( 1 + \frac{r}{\rho} \frac{d\rho}{dr} \biggr) A_\ell^m - \ell(\ell+1) B_\ell^m = 0.   
\label{Aprime} 
\end{equation} 
Equation (\ref{Aprime}) is the only equation that provides information regarding the star's internal structure, through the function $1 + (r/\rho)(d\rho/dr)$ involving the mass density $\rho(r)$.

Because $\delta h$ is formally of order $\Omega$, the boundary condition of Eq.~(\ref{surface_condition}) implies that $(GM/R^2) r_a \delta v^a(r=R) = O(\Omega^2)$. Inserting Eqs.~(\ref{va_decomposed1}) and (\ref{va_decomposed2}), we find that this condition implies 
\begin{equation} 
A_\ell^m(r=R) = 0. 
\label{BC_A} 
\end{equation} 
The radial component of the velocity field must therefore vanish on the stellar surface. 

\subsection{Truncation and projections} 

Insertion of Eqs.~(\ref{va_decomposed1}) and (\ref{va_decomposed2}) within Eq.~(\ref{curlEuler}) leads to an infinite sequence of equations for the radial functions $A_\ell^m(r)$, $B_\ell^m(r)$, and $C_\ell^m(r)$; the sequence is such that a function with a given value of $\ell$ is coupled to an infinite number of functions with distinct values of $\ell$. Any hope of finding a solution to the perturbation equations must be based on a truncation of the system; the truncation implies that the solution can only be approximate. Fortunately, numerical exploration reveals that except for isolated values of $\omega$ (which correspond to eigenfrequencies of subdominant Lockitch-Friedman inertial modes\footnote{The dominant modes are those listed in Sec.~\ref{sec:intro}. All other modes are subdominant.}), radial functions with the lowest values of $\ell$ are much larger than functions with larger values of $\ell$. An excellent representation of the solution can therefore be obtained with handfuls of terms. 

For $m=\pm 2$ we include the set of radial functions
\begin{equation}
\{ C_2^{\pm 2}, A_3^{\pm 2}, B_3^{\pm 2}, C_4^{\pm 2}, A_5^{\pm 2}, B_5^{\pm 2}, C_6^{\pm 2}, A_7^{\pm 2}, B_7^{\pm 2} \}
\label{set_m2} 
\end{equation}
in our truncated system of perturbation equations. For $m=\pm 1$ we include the set
\begin{equation} 
\{ A_1^{\pm 1}, B_1^{\pm 1}, C_2^{\pm 1}, A_3^{\pm 1}, B_3^{\pm 1}, C_4^{\pm 1}, A_5^{\pm 1}, B_5^{\pm 1}, C_6^{\pm 1} \}.
\label{set_m1} 
\end{equation}
And for $m=0$ we include
\begin{equation} 
\{ A_1^{0}, B_1^{0}, C_2^{0}, A_3^{0}, B_3^{0}, C_4^{0}, A_5^{0}, B_5^{0}, C_6^{0} \}.
\label{set_m0} 
\end{equation}
For $m=\pm 1$ and $m=0$ we find that $A_1^m$, $B_1^m$, and $C_2^m$ dominate the description of the solution by at least two orders of magnitude (except for isolated frequencies, as was mentioned previously). For $m=\pm 2$ we shall find that the only nonvanishing radial function is $C_2^{\pm 2}(r)$.  

To obtain an explicit listing of equations for the radial functions, we take the components of Eq.~(\ref{curlEuler}) and project them into spherical-harmonic components. We thus define 
\begin{subequations} 
\label{Zdef} 
\begin{align} 
(Z_\ell^m)^r &:= \int (Q^r - q^r) \bar{Y}_\ell^m\, d\Omega, 
\label{Zr_def} \\
(Z_\ell^m)^\theta &:= \int \sin\theta (Q^\theta - q^\theta) \bar{Y}_\ell^m\, d\Omega, 
\label{Ztheta_def} \\ 
(Z_\ell^m)^\phi &:= \int \sin^2\theta (Q^\phi - q^\phi) \bar{Y}_\ell^m\, d\Omega, 
\label{Zphi_def}
\end{align} 
\end{subequations} 
and the perturbation equations become $(Z_\ell^m)^a = 0$. As was pointed out, only two of these equations are independent. They are supplemented with Eq.~(\ref{Aprime}) and the surface condition of Eq.~(\ref{BC_A}).   

As we shall see in detail below, we find that the equations $(Z_\ell^m)^r = 0$ provide algebraic relations between the radial functions, while $(Z_\ell^m)^\theta = 0$ and $(Z_\ell^m)^\phi = 0$ give rise to equations involving the functions and their first derivatives. The general strategy to obtain a workable system of equations is to use the radial equations to obtain $C_\ell^m$ algebraically in terms of $A_\ell^m$ and $B_\ell^m$, invoke Eq.~(\ref{Aprime}) to express $dA_\ell^m/dr$ also in terms of $A_\ell^m$ and $B_\ell^m$, and finally, to solve the angular equations for $dB_\ell^m/dr$. We shall now go through these steps for each value of $m$.      

\subsection{Equations for $m=\pm 2$} 
\label{sec:eqns_m2} 

Setting $w := \omega/\Omega$, the listing of $(Z_\ell^m)^a$ for $m = 2$ is  
\begin{subequations} 
\label{Z_m2} 
\begin{align} 
(Z_2^2)^r &= -(6w - 8)\,  C_2^2 + \frac{4}{7}\sqrt{7}\, A_3^2 + \frac{16}{7}\sqrt{7}\,  B_3^2 
+ \frac{2}{5}\sqrt{30\pi}\,  (r/R)^3, \\
(Z_3^2)^\theta &= \frac{4}{3}\,  r \frac{dA_3^2}{dr} - \frac{4}{33}\sqrt{11}\,  r \frac{d A_5^2}{dr} 
+ 2(w-2)\,  r \frac{d B_3^2}{dr} - \frac{8}{11}\sqrt{11}\,  r \frac{d B_5^2}{dr} 
- \frac{2}{7}\sqrt{7} w\,  r \frac{d C_2^2}{dr} + \frac{2}{21}\sqrt{21} (5w-14)\,  r \frac{d C_4^2}{dr} 
\nonumber \\ & \quad \mbox{}
+ \frac{8}{7}\sqrt{7}\,  C_2^2 - (2w-4)\,  A_3^2 - 8\,  B_3^2 - \frac{40}{21}\sqrt{21}\,  C_4^2 
+ \frac{2}{35}\sqrt{210\pi}\,  (r/R)^3, \\
(Z_4^2)^r &= -\frac{20}{21}\sqrt{21}\,  A_3^2 + \frac{20}{7}\sqrt{21}\,  B_3^2 - (20w-36)\,  C_4^2 
+ \frac{8}{33}\sqrt{231}\, A_5^2 + \frac{16}{11}\sqrt{231}\, B_5^2, \\
(Z_5^2)^\theta &= -\frac{4}{33}\sqrt{11}\, r \frac{d A_3^2}{dr} + \frac{44}{39}\, r \frac{dA_5^2}{dr} 
- \frac{8}{143}\sqrt{66}\, r \frac{d A_7^2}{dr}  + \frac{4}{11}\sqrt{11}\, r \frac{d B_3^2}{dr} 
+ \frac{2}{13}(13w-28) \, r \frac{d B_5^2}{dr} - \frac{64}{143}\sqrt{66}\, r \frac{d B_7^2}{dr} 
\nonumber \\ & \quad \mbox{}
- \frac{4}{33}\sqrt{231} (w-1) \, r \frac{dC_4^2}{dr} + \frac{4}{143}\sqrt{286} (7w-18) \, r \frac{d C_6^2}{dr} 
+ \frac{16}{33}\sqrt{231}\, C_4^2 - (2w-4) \, A_5^2 - \frac{112}{143}\sqrt{286}\, C_6^2 - 8 \, B_5^2, \\ 
(Z_6^2)^r &= -\frac{56}{143}\sqrt{286}\, A_5^2 + \frac{280}{143}\sqrt{286}\, B_5^2 - (42w-80) \, C_6^2 
+ \frac{12}{13}\sqrt{39}\, A_7^2 + \frac{96}{13}\sqrt{39}\, B_7^2, \\ 
(Z_7^2)^\theta &= -\frac{8}{143}\sqrt{66}\, r \frac{dA_5^2}{dr} + \frac{236}{221}\, r \frac{dA_7^2}{dr} 
+ \frac{40}{143}\sqrt{66}\, r \frac{dB_5^2}{dr} + \frac{2}{221} (221 w - 486) \, r \frac{dB_7^2}{dr} 
- \frac{2}{13}\sqrt{39} (3w-4) \, r \frac{d C_6^2}{dr} 
\nonumber \\ & \quad \mbox{}
+ \frac{24}{13}\sqrt{39}\, C_6^2 - (2w-4) \, A_7^2 - 8 \, B_7^2. 
\end{align} 
\end{subequations} 
The expressions for $m=-2$ can be obtained from these by applying the identities of Eq.~(\ref{ABCidentities1}). 

The equations $(Z_\ell^2)^r = 0$ for $\ell = 2, 4, 6$ provide algebraic solutions for $C_\ell^2$ in terms of $A_\ell^2$ and $B_\ell^2$. Making the substitutions in $(Z_\ell^2)^\theta = 0$ for $\ell = 3, 5, 7$, and making use of Eq.~(\ref{Aprime}) to eliminate derivatives of $A_\ell^2$, we obtain for $B_\ell^2$ a system of equations of the schematic form 
\begin{equation} 
r \frac{dB_\ell^2}{dr} = \sum_{\ell' = 3, 5, 7} \bigl( {\cal P}_\ell^{\ell'}\, A_{\ell'}^2 
+ {\cal Q}_\ell^{\ell'}\, B_{\ell'}^2 \bigr) + F_\ell^2, 
\label{Bprime_m2} 
\end{equation} 
where ${\cal P}_\ell^{\ell'}$ and ${\cal Q}_\ell^{\ell'}$ are coefficients that depend on $w$ and $r$, and $F_\ell^2 \propto (r/R)^3$ are driving terms that originate from $q^a$, the curl of the tidal force density. The radial functions are therefore determined by the system of differential equations provided by Eqs.~(\ref{Aprime}) and (\ref{Bprime_m2}), the algebraic equations for $C_\ell^2$, and the boundary conditions of Eq.~(\ref{surface_condition}). 

\subsection{Equations for $m=\pm 1$} 
\label{sec:eqns_m1} 

The listing of $(Z_\ell^m)^a$ for $m = 1$ is  
\begin{subequations} 
\label{Z_m1} 
\begin{align} 
(Z_1^1)^\theta &= -\frac{8}{5}\, r \frac{dA_1^1}{dr} + \frac{4}{35}\sqrt{14}\, r \frac{dA_3^1}{dr} 
- \frac{1}{5}(5w - 3)\, r \frac{dB_1^1}{dr} + \frac{16}{35} \sqrt{14}\, r \frac{dB_3^1}{dr} 
- \frac{1}{5} \sqrt{5} (3w-5)\, r \frac{dC_2^1}{dr} 
\nonumber \\ & \quad \mbox{}
+ (w-1)\, A_1^1 + 2\, B_1^1 + \frac{6}{5}\sqrt{5}\, C_2^1 - \frac{3}{5}\sqrt{6\pi}\, (r/R)^3, \\ 
(Z_2^1)^r &= \frac{6}{5}\sqrt{5}\, A_1^1 - \frac{6}{5}\sqrt{5}\, B_1^1 + (6w-4)\, C_2^1 
- \frac{8}{35}\sqrt{70}\, A_3^1 - \frac{32}{35}\sqrt{70}\, B_3^1 + \frac{2}{5}\sqrt{30\pi} (r/R)^3, \\ 
(Z_3^1)^\theta &= \frac{4}{35}\sqrt{14}\, r \frac{dA_1^1}{dr} - \frac{16}{15}\, r \frac{d A_3^1}{dr} 
+ \frac{4}{231}\sqrt{770}\, r \frac{dA_5^1}{dr} - \frac{4}{35}\sqrt{14}\, r \frac{d B_1^1}{dr} 
- \frac{1}{5}(5w-7)\, r \frac{d B_3^1}{dr} + \frac{8}{77}\sqrt{770}\, r \frac{dB_5^1}{dr} 
\nonumber \\ & \quad \mbox{}
+ \frac{4}{35}\sqrt{70}w\, r \frac{d C_2^1}{dr} - \frac{1}{21}\sqrt{105}(5w-7)\, r \frac{d C_4^1}{dr} 
- \frac{8}{35}\sqrt{70}\, C_2^1 + (w-1)\, A_3^1 + 2\, B_3^1 + \frac{10}{21}\sqrt{105}\, C_4^1 
\nonumber \\ & \quad \mbox{}
+ \frac{8}{35}\sqrt{21\pi} (r/R)^3, \\ 
(Z_4^1)^r &= \frac{10}{21}\sqrt{105}\, A_3^1 - \frac{10}{7}\sqrt{105}\, B_3^1 + (20w-18)\, C_4^1 
- \frac{16}{33}\sqrt{66}\, A_5^1 - \frac{32}{11}\sqrt{66}\, B_5^1, \\ 
(Z_5^1)^\theta &= \frac{4}{231}\sqrt{770}\, r \frac{d A_3^1}{dr} - \frac{40}{39}\, r \frac{d A_5^1}{dr} 
- \frac{4}{77}\sqrt{770}\, r \frac{d B_3^1}{dr} - \frac{1}{13} (13w - 19)\, r \frac{d B_5^1}{dr} 
+ \frac{4}{33}\sqrt{66}(2w-1)\, r \frac{d C_4^1}{dr} 
\nonumber \\ & \quad \mbox{}
- \frac{1}{143}\sqrt{5005}(7w-9)\, r \frac{d C_6^1}{dr} 
- \frac{16}{33}\sqrt{66}\, C_4^1 + (w-1)\, A_5^1 + 2\, B_5^1 + \frac{14}{143}\sqrt{5005}\, C_6^1, \\ 
(Z_6^1)^r &= \frac{14}{143}\sqrt{5005}\, A_5^1 - \frac{70}{143}\sqrt{5005}\, B_5^1 
+ (42w-40)\, C_6^1. 
\end{align} 
\end{subequations} 
Expressions for $m=-1$ are obtained from these by making use of Eq.~(\ref{ABCidentities2}). 

The equations $(Z_\ell^1)^r = 0$ for $\ell = 2, 4, 6$ determine $C_\ell^1$ in terms of $A_\ell^1$ and $B_\ell^1$. Substituting these in $(Z_\ell^1)^\theta = 0$ for $\ell = 1, 3, 5$, and eliminating derivatives of $A_\ell^1$ with Eq.~(\ref{Aprime}), we obtain for $B_\ell^1$ a system of equations of the schematic form 
\begin{equation} 
r \frac{dB_\ell^1}{dr} = \sum_{\ell' = 1, 3, 5} \bigl( {\cal P}_\ell^{\ell'}\, A_{\ell'}^1 
+ {\cal Q}_\ell^{\ell'}\, B_{\ell'}^1 \bigr) + F_\ell^1, 
\label{Bprime_m1} 
\end{equation} 
where ${\cal P}_\ell^{\ell'}$ and ${\cal Q}_\ell^{\ell'}$ are coefficients --- distinct from those appearing in Eq.~(\ref{Bprime_m2}) --- that depend on $w$ and $r$, and $F_\ell^1 \propto (r/R)^3$ are driving terms that originate from $q^a$. The radial functions are therefore determined by the system of differential equations provided by Eqs.~(\ref{Aprime}) and (\ref{Bprime_m1}), the algebraic equations for $C_\ell^1$, and the boundary conditions of Eq.~(\ref{surface_condition}).    

\subsection{Equations for $m=0$} 
\label{sec:eqns_m0} 

For $m=0$ we have 
\begin{subequations} 
\label{Z_m0} 
\begin{align} 
(Z_2^0)^r &= \frac{4}{5}\sqrt{15}\, A_1^0 - \frac{4}{5}\sqrt{15}\, B_1^0 - 6w\, C_2^0 
- \frac{12}{35}\sqrt{35}\, A_3^0 - \frac{48}{35}\sqrt{35}\, B_3^0 - \frac{12}{5}\sqrt{5\pi}\, (r/R)^3, \\ 
(Z_2^0)^\phi &= -\frac{2}{15}\sqrt{15}w\, r \frac{dB_1^0}{dr} + \frac{12}{35}\sqrt{35}w\, r \frac{dB_3^0}{dr} 
+ \frac{4}{7}\, r \frac{dC_2^0}{dr} + \frac{8}{7}\sqrt{5}\, r \frac{d C_4^0}{dr} 
\nonumber \\ & \quad \mbox{}
+ \frac{2}{15}\sqrt{15}w\, A_1^0 - \frac{40}{7}\, C_2^0 - \frac{12}{35}\sqrt{35}w\, A_3^0 
+ \frac{32}{7}\sqrt{5}\, C_4^0, \\ 
(Z_4^0)^r &= \frac{40}{21}\sqrt{7}\, A_3^0 - \frac{40}{7}\sqrt{7}\, B_3^0 - 20w\, C_4^0 
- \frac{40}{33}\sqrt{11}\, A_5^0 - \frac{80}{11}\sqrt{11}\, B_5^0, \\ 
(Z_4^0)^\phi &= -\frac{4}{7}\sqrt{7}w\, r \frac{dB_3^0}{dr} + \frac{10}{11}\sqrt{11}w\, r \frac{dB_5^0}{dr} 
- \frac{16}{35}\sqrt{5}\, r \frac{dC_2^0}{dr} + \frac{40}{77}\, r \frac{d C_4^0}{dr} 
+ \frac{140}{143}\sqrt{13}\, r \frac{dC_6^0}{dr} 
\nonumber \\ & \quad \mbox{}
+ \frac{48}{35}\sqrt{5}\, C_2^0 + \frac{4}{7}\sqrt{7}w\, A_3^0 
- \frac{1520}{77}\, C_4^0 - \frac{10}{11}\sqrt{11}w\, A_5^0 + \frac{840}{143}\sqrt{13}\, C_6^0, \\ 
(Z_6^0)^r &=  \frac{84}{143}\sqrt{143}\, A_5^0 - \frac{420}{143}\sqrt{143}\, B_5^0 - 42w\, C_6^0, \\ 
(Z_6^0)^\phi &= -\frac{30}{143}\sqrt{143}w\, r\frac{dB_5^0}{dr} - \frac{80}{143}\sqrt{13}\, r \frac{dC_4^0}{dr} 
+ \frac{28}{55}\, r \frac{d C_6^0}{dr} + \frac{400}{143}\sqrt{13}\, C_4^0 + \frac{30}{143}\sqrt{143}w\, A_5^0 
- \frac{2296}{55}\, C_6^0. 
\end{align} 
\end{subequations} 
It might be noted that in this case, the independent equations are provided by $(Z_\ell^0)^r$ and $(Z_\ell^0)^\phi$, with $\ell = 2, 4, 6$. For $m = \pm 2$ and $\pm 1$, the independent equations came from $(Z_\ell^m)^r$ and $(Z_\ell^m)^\theta$, with both even and odd values of $\ell$.  

The equations $(Z_\ell^0)^r = 0$ for $\ell = 2, 4, 6$ give $C_\ell^0$ algebraically in terms of $A_\ell^0$ and $B_\ell^0$. Substituting these relations in $(Z_\ell^0)^\phi = 0$ for $\ell = 2, 4, 6$, and using Eq.~(\ref{Aprime}) to replace the derivatives of $A_\ell^0$, we obtain for $B_\ell^0$ the system of equations 
\begin{equation} 
r \frac{dB_\ell^0}{dr} = \sum_{\ell' = 1, 3, 5} \bigl( {\cal P}_\ell^{\ell'}\, A_{\ell'}^0 
+ {\cal Q}_\ell^{\ell'}\, B_{\ell'}^0 \bigr) + F_\ell^0, 
\label{Bprime_m0} 
\end{equation} 
where ${\cal P}_\ell^{\ell'}$ and ${\cal Q}_\ell^{\ell'}$ are again coefficients that depend on $w$ and $r$, distinct from those appearing in Eqs.~(\ref{Bprime_m2}) and (\ref{Bprime_m1}), and $F_\ell^0 \propto (r/R)^3$ are driving terms. The radial functions are therefore determined by the system of differential equations provided by Eqs.~(\ref{Aprime}) and (\ref{Bprime_m0}), the algebraic equations for $C_\ell^0$, and the boundary conditions of Eq.~(\ref{surface_condition}).    

\section{Solution to the perturbation equations} 
\label{sec:solution} 

In this section we shall find solutions to the perturbation equations developed in Sec.~\ref{sec:perturbation}. For $m=\pm 2$ the solution will be exact, and it will be obtained analytically. For $m=\pm 1$ and $m=0$ the solutions will be approximate, and obtained numerically. 

\subsection{Numerical methods} 

For $m=\pm 1$ and $m = 0$ we have the set of variables displayed in Eqs.~(\ref{set_m1}) and (\ref{set_m0}), the system of first-order differential equations provided by Eqs.~(\ref{Aprime}), (\ref{Bprime_m1}), and (\ref{Bprime_m0}), and the surface conditions of Eq.~(\ref{BC_A}). The various coefficients ${\cal P}_\ell^{\ell'}$ and ${\cal Q}_\ell^{\ell'}$ depend on $w := \omega/\Omega$, the dimensionless frequency of the gravitomagnetic tidal field, and they depend on $r$ through the function $1 + (r/\rho) (d\rho/dr)$, which encodes the relevant details of the star's internal structure. For concreteness and simplicity in this section, we adopt for the stellar fluid a polytropic equation of state $p = K \rho^2$, where $K$ is a constant. For this model we have that $K = 2GR^2/\pi$ and 
\begin{equation} 
\rho = \frac{M}{4R^2 r} \sin(\pi r /R),   
\label{density} 
\end{equation} 
from which it follows that 
\begin{equation} 
1 + \frac{r}{\rho} \frac{d\rho}{dr} = (\pi r/R) \cot(\pi r/R).   
\end{equation} 
This density model corresponds to $k = 1$ in Eq.~(\ref{rho_kmodel_intro}). 

A local analysis of the differential equations (\ref{Aprime}), (\ref{Bprime_m1}), and (\ref{Bprime_m0}) near $r=0$ reveals that for $m = \pm 1$ and $m=0$, $A_\ell^m$ and $B_\ell^m$ admit the expansions 
\begin{equation} 
A_\ell^m = \sum_{k=\ell}^\infty a_\ell^k (r/R)^k, \qquad 
B_\ell^m = \sum_{k=\ell}^\infty b_\ell^k (r/R)^k,
\end{equation} 
where $a_\ell^k$ and $b_\ell^k$ are constant coefficients. The analysis reveals also that $a_1^1$, $a_3^3$, and $a_5^5$ can be specified freely, and that all remaining coefficients are determined by the differential equations. 

A local analysis of the differential equations near $r=R$, which takes into account the boundary conditions of Eq.~(\ref{BC_A}), proceeds on the basis of the expansions 
\begin{equation} 
A_\ell^m = \sum_{k=1}^\infty \alpha_\ell^k (r/R - 1)^k, \qquad 
B_\ell^m = \sum_{k=0}^\infty \beta_\ell^k (r/R - 1)^k,
\end{equation} 
where $\alpha_\ell^k$ and $\beta_\ell^k$ are constant coefficients. The analysis shows that $\alpha_1^1$, $\alpha_3^1$, and $\alpha_5^1$ can be specified freely, and that all remaining coefficients are determined by the differential equations. 

To find the solutions to Eqs.~(\ref{Aprime}), (\ref{Bprime_m1}), and (\ref{Bprime_m0}) we construct an inner solution in the interval $0 \leq r \leq r^\sharp$, an outer solution in the interval $r^\sharp \leq r \leq R$, and demand that the inner and outer solutions agree at $r = r^\sharp$, with $r^\sharp$ denoting an arbitrary middle point. To describe the method we collectively denote by $\bm{e}$ the set of radial functions $A_\ell^m$ and $B_\ell^m$. For the inner solution we construct a particular solution $\bm{e}^{\rm in}_{\rm part}$ to the differential equations by making a random selection of the constants $\{ a_1^1, a_3^3, a_5^5 \}$. We also form a basis of functions $\bm{e}^{\rm in}_n$ (with $n = \{1, 2, 3\}$), defined to be solutions to the homogeneous version of the differential equations, with all driving terms $F_\ell^m$ switched off; each member of the basis is constructed with a different random selection of the constants $\{ a_1^1, a_3^3, a_5^5 \}$. The correct inner solution to the differential equations is 
\begin{equation} 
\bm{e}^{\rm in} = \bm{e}^{\rm in}_{\rm part} + p_1\, \bm{e}^{\rm in}_1 + p_2\, \bm{e}^{\rm in}_2
+ p_3\, \bm{e}^{\rm in}_3, 
\end{equation} 
where $\{ p_1, p_2, p_3 \}$ are unknown coefficients. Moving on to the outer solution, we construct a particular solution $\bm{e}^{\rm out}_{\rm part}$ to the differential equations by making a random selection of the constants $\{ \alpha_1^1, \alpha_3^1, \alpha_5^1 \}$. The functions $\bm{e}^{\rm out}_n$ are solutions to the homogeneous version of the differential equations, with each member associated with a different random selection of the constants $\{ \alpha_1^1, \alpha_3^1, \alpha_5^1 \}$. The correct outer solution to the differential equations is
\begin{equation} 
\bm{e}^{\rm out} = \bm{e}^{\rm out}_{\rm part} + q_1\, \bm{e}^{\rm out}_1 + q_2\, \bm{e}^{\rm out}_2
+ q_3\, \bm{e}^{\rm out}_3, 
\end{equation} 
where $\{ q_1, q_2, q_3 \}$ are unknown coefficients. The constants $\{ p_n, q_n \}$ are determined by the requirement that the inner and outer solutions agree at $r = r^\sharp$:  
\begin{equation} 
\bm{e}^{\rm in}(r=r^\sharp) = \bm{e}^{\rm out}(r=r^\sharp). 
\end{equation} 
This gives us six equations for six unknowns, and we finally have our solution in the complete interval $0 \leq r \leq R$. 

\subsection{Solution for $m=\pm 2$} 

We observe from the explicit form of Eq.~(\ref{Bprime_m2}) that $F_\ell^{\pm 2} = 0$; there is no driving term for the radial functions $A_\ell^{\pm 2}$ and $B_\ell^{\pm 2}$. This implies that they may be set equal to zero, and we find that the remaining perturbation equations produce 
\begin{equation} 
C_2^{\pm 2} = \pm \frac{1}{5}\sqrt{30\pi}\, \frac{(r/R)^3}{3w \mp 4}, 
\label{C22_solution} 
\end{equation} 
together with $C_4^{\pm 2} = C_6^{\pm 2} = 0$. While this provides a solution to our truncated system of equations,  involving only the set of Eq.~(\ref{set_m2}), it was verified that  
\begin{equation} 
\delta v^a = \mp \frac{i}{5}\sqrt{30\pi}\, \frac{{\cal B}_0}{c^2} R^3 \sin\iota(\cos\iota \pm 1)\,
\frac{2 \mp w}{3w \mp 4}\, (r/R)^3\, (X_2^{\pm 2})^a\, e^{-i\omega t} + \mbox{cc} 
\label{dv_m2} 
\end{equation} 
is an {\it exact solution} to the perturbation equations for $m = \pm 2$; we recall that $w := \omega/\Omega$. We notice that the solution is formally infinite when $w = \pm 4/3$. These frequencies coincide with the eigenfrequencies of $r$-modes for a barotropic star. The precise connection with $r$-modes will be made below in Sec.~\ref{sec:oscillator}.  

\subsection{Solution for $m=\pm 1$ and $m=0$} 

\begin{figure} 
\includegraphics[width=0.7\linewidth]{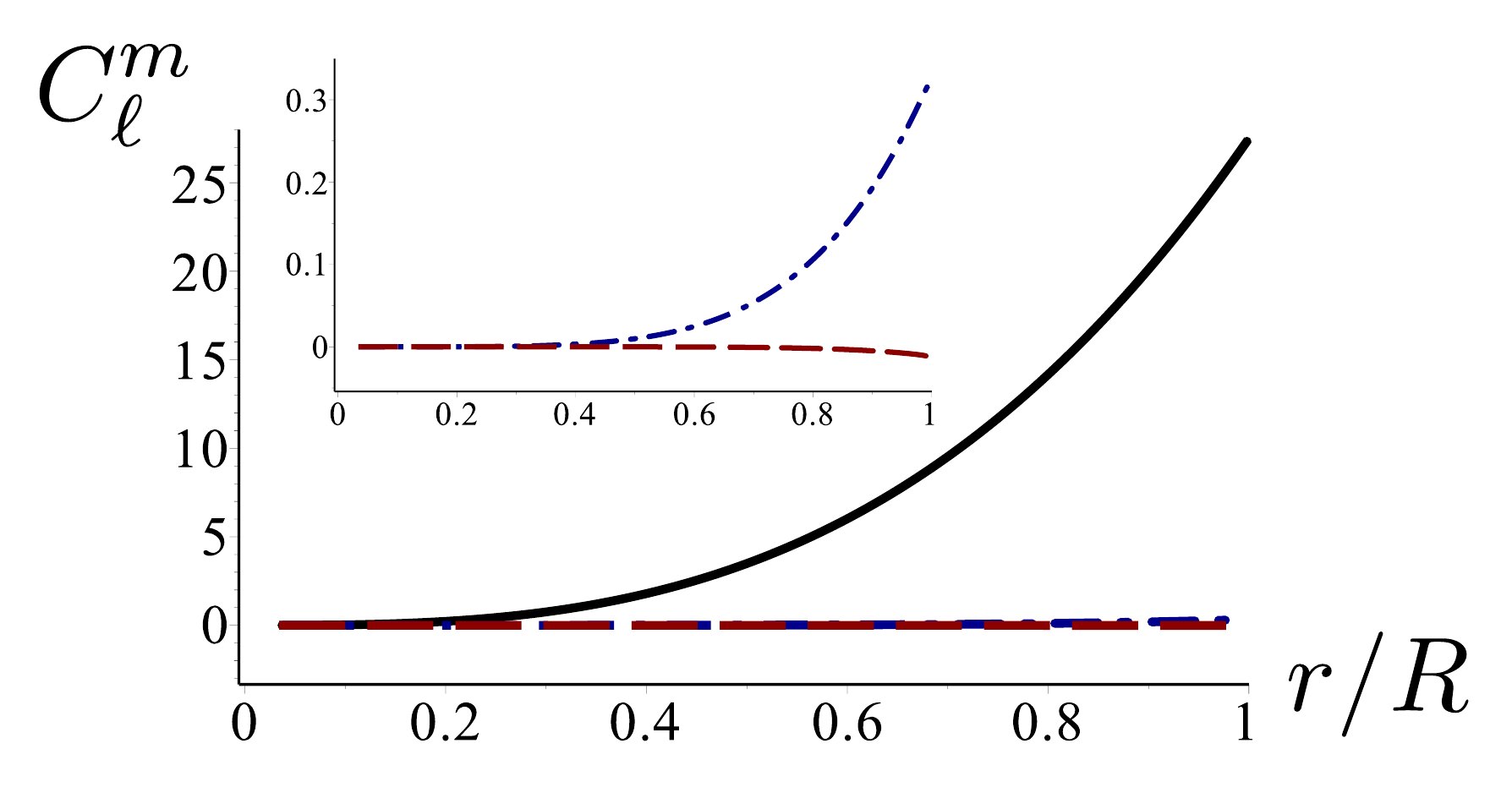}
\caption{Radial functions $C^m_\ell(r)$ for $m = 1$, computed for $w = 1.0$. The solid black line corresponds to $\ell = 2$, the dash-dotted blue line corresponds to $\ell=4$, and the dashed red line corresponds to $\ell = 6$.}   
\label{fig:Cm1} 
\end{figure} 

\begin{figure} 
\includegraphics[width=0.7\linewidth]{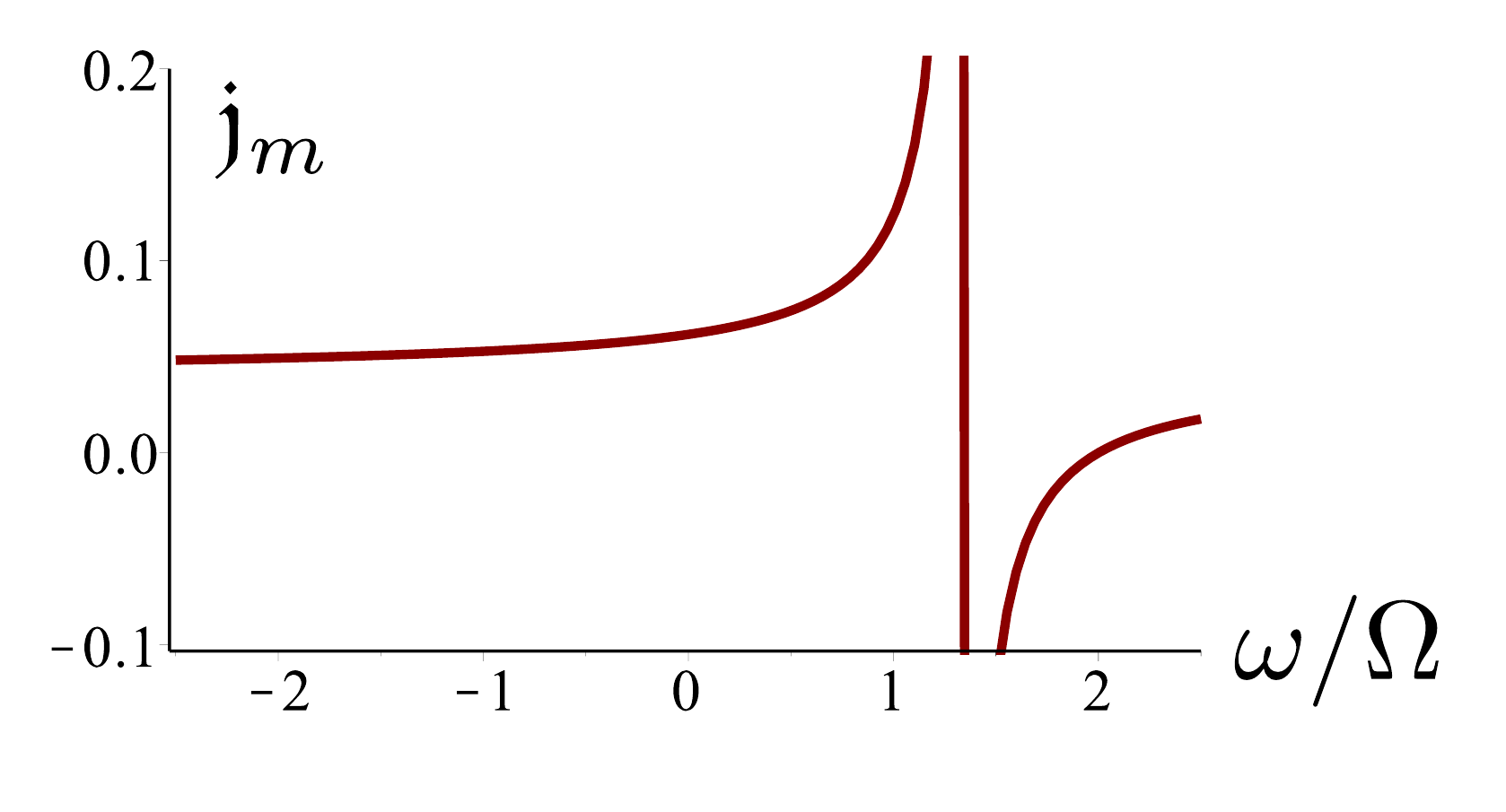}
\caption{Radial functions $C^m_\ell(r)$ for $m = 0$, computed for $w = 1.0$. The solid black line corresponds to $\ell = 2$, the dash-dotted blue line corresponds to $\ell=4$, and the dashed red line corresponds to $\ell = 6$.} 
\label{fig:Cm0} 
\end{figure} 

Next we present a very small sampling of our numerical results for $m = \pm 1$ and $m = 0$. In Fig.~\ref{fig:Cm1} we display the functions $C_\ell^1(r)$ for $\ell = 2, 4, 6$, computed for $w = 1.0$. In Fig.~\ref{fig:Cm0} we show $C_\ell^0(r)$ for $\ell = 2, 4, 6$, also computed for $w = 1.0$. As advertised, in both cases $|C_4^m|$ is smaller than $C_2^m$ by about two orders of magnitude, and $|C_6^m|$ is smaller still. This behavior is shared by the other radial functions: we have that $|A_5^m| < |A_3^m| \ll |A_1^m|$ and $|B_5^m| < |B_3^m| \ll |B_1^m|$. As was pointed out, this hierarchy is satisfied for all values of $w$, except for eigenfrequencies of subdominant inertial modes; the hierarchy continues to be respected when $w$ is an eigenfrequency of a dominant mode.   

\subsection{Current quadrupole moment} 
\label{sec:quadmoment}  

A meaningful measure of the velocity perturbation is provided by the current quadrupole moment $J^{jk}$, defined by 
\begin{equation} 
J^{jk} = \int \epsilon^{(j}_{\ \, pq} x^{k)} x^p \rho\, \delta v^q\, dV, 
\label{J_def} 
\end{equation} 
where $\delta v^q$ is the velocity field of Eqs.~(\ref{va_decomposed1}) and (\ref{va_decomposed2}), and $dV$ is the volume element. As in Sec.~\ref{sec:gravitomagnetic}, we use indices $j,k,p,\ldots$ in all equations formulated in Cartesian coordinates. 

To calculate $J^{jk}$ we rely on the identities 
\begin{subequations} 
\label{iden} 
\begin{align} 
\epsilon^{(j}_{\ pq} r^{k)} r^p\, r^q\, Y_\ell^m &= 0, 
\label{iden1} \\ 
\int \epsilon^{(j}_{\ pq} r^{k)} r^p\, (Y_\ell^m)^q\, d\Omega &= 0, 
\label{iden2} \\ 
\int \epsilon^{(j}_{\ pq} r^{k)} r^p\, (X_\ell^m)^q\, d\Omega&= 
\frac{8\pi}{5 r}\, (\bar{\scrpt Y}_2^m)^{jk}\, \delta_{\ell,2} 
\label{iden3} 
\end{align} 
\end{subequations} 
involving the radial, polar, and axial harmonics of Sec.~\ref{subsec:harmonics}, as well as the unit radial vector $r^j = x^j/r$; in Eq.~(\ref{iden3}) we have the constant tensors (see Box~1.5 of Ref.~\cite{poisson-will:14}) 
\begin{equation} 
(\bar{\scrpt Y}_2^{\pm 2})^{jk} = \frac{1}{8} \sqrt{\frac{30}{\pi}} \left( 
\begin{array}{ccc} 
1 & \pm i & 0 \\ 
\pm i & -1 & 0 \\ 
0 & 0 & 0 
\end{array} \right), \qquad 
(\bar{\scrpt Y}_2^{\pm 1})^{jk} = \frac{1}{8} \sqrt{\frac{30}{\pi}} \left( 
\begin{array}{ccc} 
0 & 0 & \mp 1 \\ 
0 & 0 & -i \\ 
\mp 1 & -i & 0 
\end{array} \right), \qquad 
(\bar{\scrpt Y}_2^{0})^{jk} = \frac{1}{4} \sqrt{\frac{5}{\pi}} \left( 
\begin{array}{ccc} 
-1 & 0 & 0 \\ 
0 & -1 & 0 \\ 
0 & 0 & 2 
\end{array} \right), 
\label{STFtensors} 
\end{equation} 
which are all symmetric and tracefree. 

The result of Eq.~(\ref{iden1}) follows simply from the antisymmetry of $\epsilon^j_{\ pq}$ with respect to the indices $p$ and $q$ and the symmetry of $r^p r^q$; the presence of the spherical harmonics is irrelevant. The identity of Eq.~(\ref{iden2}) is derived by importing the definition of the polar harmonics, $(Y_\ell^m)^q = \nabla^q Y_\ell^m$, and performing an integration by parts; symmetrization of the $jk$ indices ensures that the result vanishes. To establish the third equation we insert the definition of the axial harmonics, $(X_\ell^m)^q = \epsilon^{qrs} \nabla_r Y_\ell^m r_s$, combine the permutation symbols, and perform an integration by parts to obtain 
\begin{equation} 
\int \epsilon^{(j}_{\ pq} r^{k)} r^p\, (X_\ell^m)^q\, d\Omega 
= \frac{3}{r} \int (r^j r^k - \tfrac{1}{3} \delta^{jk}) Y_\ell^m\, d\Omega.
\end{equation} 
In the next step we rely on the fact that with the help of the constant tensors introduced in Eq.~(\ref{STFtensors}), the symmetric, tracefree tensor within the integral can be decomposed in $\ell=2$ spherical harmonics, according to [Eq.~(1.164) of Ref.~\cite{poisson-will:14}] 
\begin{equation} 
r^j r^k - \frac{1}{3} \delta^{jk} = \frac{8\pi}{15} \sum_{m=-2}^2  
(\bar{\scrpt Y}_2^m)^{jk}\, \bar{Y}_2^m(\theta,\phi). 
\end{equation} 
For the final step we integrate over the angles and make use of the orthonormality of spherical harmonics.  

\begin{figure} 
\includegraphics[width=0.7\linewidth]{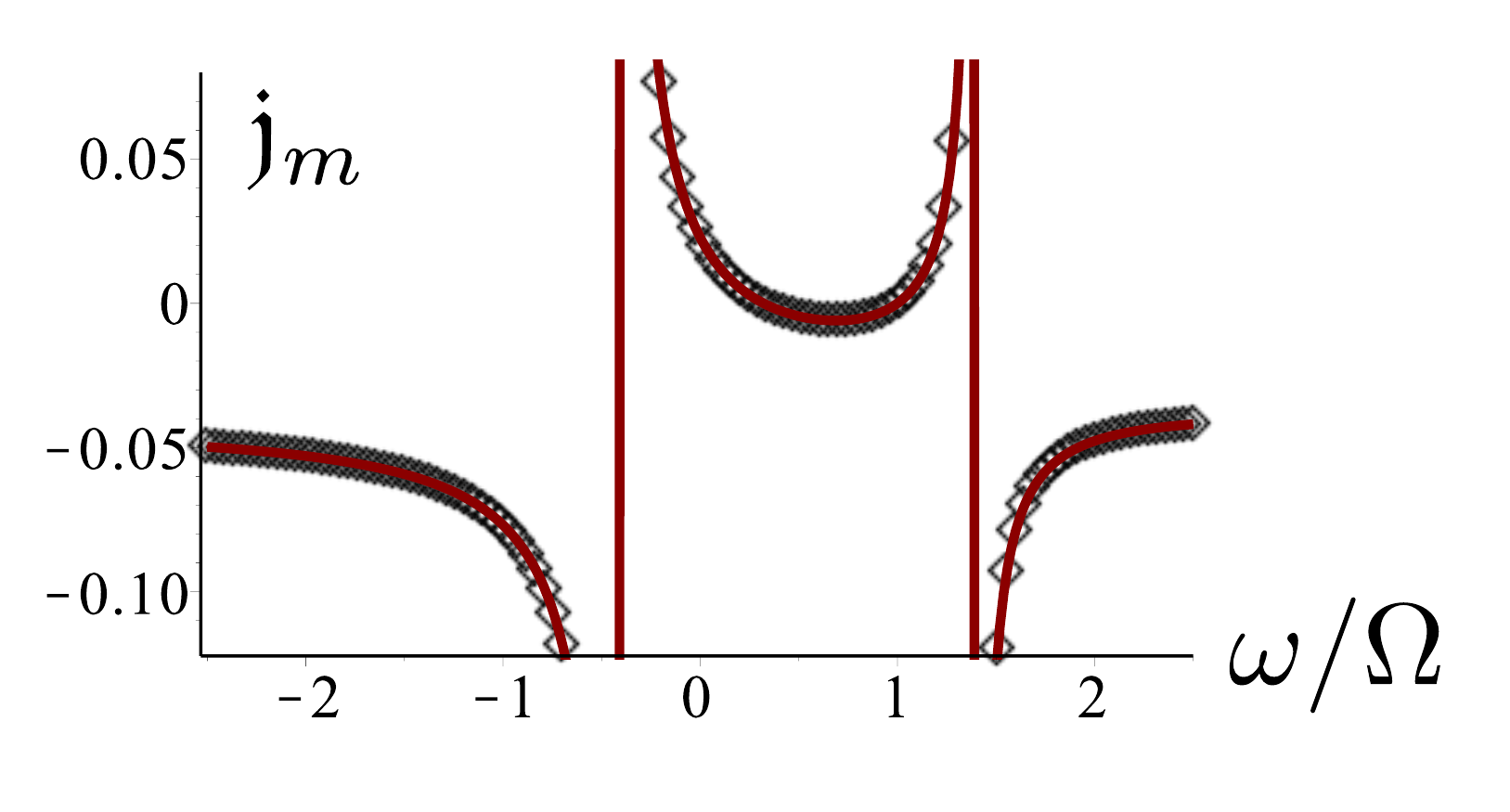}
\caption{Plot of $\gothj_m$ for $m=2$ as a function of $w := \omega/\Omega$. The vertical line at $w = 4/3 \simeq 1.3333$ corresponds to the eigenfrequency of an $r$-mode.} 
\label{fig:j_m2} 
\end{figure} 

\begin{figure} 
\includegraphics[width=0.7\linewidth]{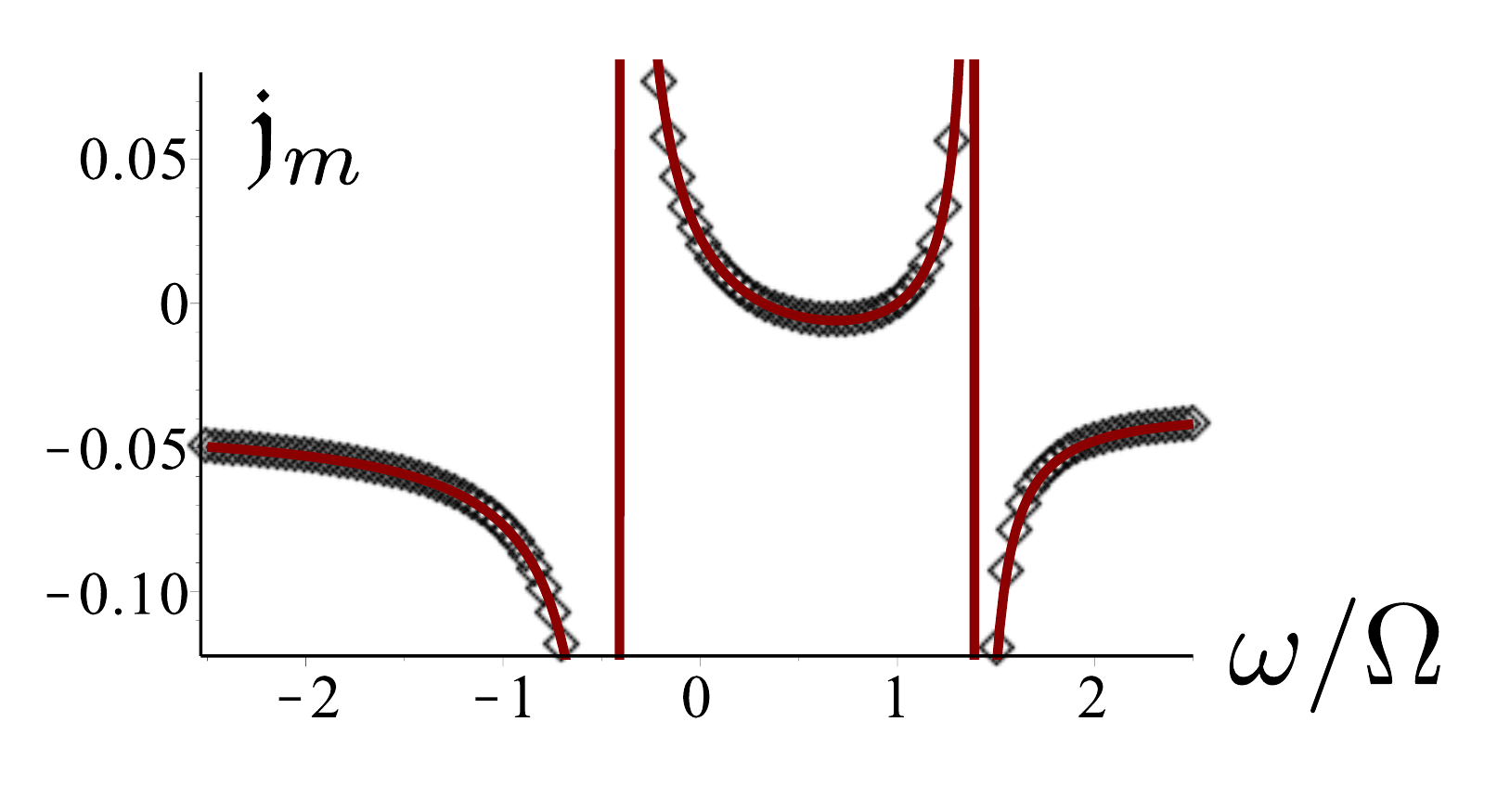}
\caption{Plot of $\gothj_m$ for $m=1$ as a function of $w$. The black diamonds result from the numerical evaluation of Eq.~(\ref{j_components}). The solid red curve is the mode-sum representation constructed in Sec.~\ref{sec:oscillator}. The vertical lines at $w \simeq -0.4130$ and $w \simeq 1.4014$ correspond to the eigenfrequencies of dominant inertial modes (polar-led modes with $m=1$ and $\ell_0 = 2$, calculated for a $p = K\rho^2$ polytrope).}   
\label{fig:j_m1} 
\end{figure} 

The identities of Eqs.~(\ref{iden}) imply that when Eqs.~(\ref{va_decomposed1}) and (\ref{va_decomposed2}) are inserted within Eq.~(\ref{J_def}), the volume integral reduces to a radial integration involving the mass density $\rho$, the functions $C_2^m(r)$, and powers of $r$. For $m = \pm 2$ we can make further use of the exact solution of Eq.~(\ref{C22_solution}). After working through the details, we arrive at 
\begin{equation} 
J^{jk} = \sum_{m=-2}^2 J^{jk}_m,  
\label{J_decomposed} 
\end{equation} 
where 
\begin{subequations} 
\label{J_components} 
\begin{align} 
J^{jk}_{\pm 2} &= \frac{{\cal B}_0}{c^2} M R^4 \sin\iota (\cos\iota \pm 1)\, \gothj_{\pm 2}(\omega) 
\left( \begin{array}{ccc} 
\sin\omega t & \mp \cos\omega t & 0 \\ 
\mp \cos\omega t & -\sin\omega t & 0 \\ 
0 & 0 & 0 
\end{array} \right), \\ 
J^{jk}_{\pm 1} &= -\frac{{\cal B}_0}{c^2} M R^4 (\cos\iota \pm 1)(2\cos\iota \mp 1)\, 
\gothj_{\pm 1}(\omega) 
\left( \begin{array}{ccc} 
0 & 0 & \cos\omega t \\ 
0 & 0 & \pm \sin\omega t \\ 
\cos\omega t & \pm \sin\omega t & 0 
\end{array} \right), \\ 
J^{jk}_{0} &= \frac{{\cal B}_0}{c^2} M R^4 \sin\iota \cos\iota\, \gothj_{0}(\omega) 
\left( \begin{array}{ccc} 
\sin\omega t & 0 & 0 \\ 
0 & \sin\omega t & 0 \\ 
0 & 0 & -2\sin\omega t 
\end{array} \right), 
\end{align} 
\end{subequations} 
with 
\begin{subequations} 
\label{j_components} 
\begin{align} 
\gothj_{\pm 2}(\omega) &:= \frac{9}{5} \frac{2 \mp w}{4 \mp 3w} 
\int_0^1 \hat{\rho}\, \hat{r}^6\, d\hat{r}, 
\label{j_components_m2} \\  
\gothj_{\pm 1}(\omega) &:= -3 \sqrt{\frac{3}{10\pi}} (1 \mp w) 
\int_0^1 \hat{\rho}\, C_2^{\pm 1}\, \hat{r}^3\, d\hat{r}, \\ 
\gothj_0(\omega) &:= \frac{3}{\sqrt{5\pi}} w 
\int_0^1 \hat{\rho}\, C_2^{0}\, \hat{r}^3\, d\hat{r}. 
\end{align} 
\end{subequations} 
We introduced $\hat{\rho} := (4\pi R^3/3M) \rho$ as a dimensionless density, and $\hat{r} := r/R$ as a dimensionless radial coordinate. By virtue of the definitions and Eqs.~(\ref{ABCidentities1}), (\ref{ABCidentities2}), and (\ref{ABCidentities3}), we have that 
\begin{equation} 
\gothj_{-2}(\omega) = \gothj_{2}(-\omega), \qquad 
\gothj_{-1}(\omega) = -\gothj_{1}(-\omega) \qquad 
\gothj_0(\omega) = \gothj_0(-\omega).  
\label{j_identities} 
\end{equation} 
A minus sign was incorporated in the definition of $\gothj_{\pm 1}$ to enforce a convention that each  $\gothj_m(\omega)$ for $m \geq 0$ is positive in a neighborhood of $\omega = 0$. 

\begin{figure} 
\includegraphics[width=0.7\linewidth]{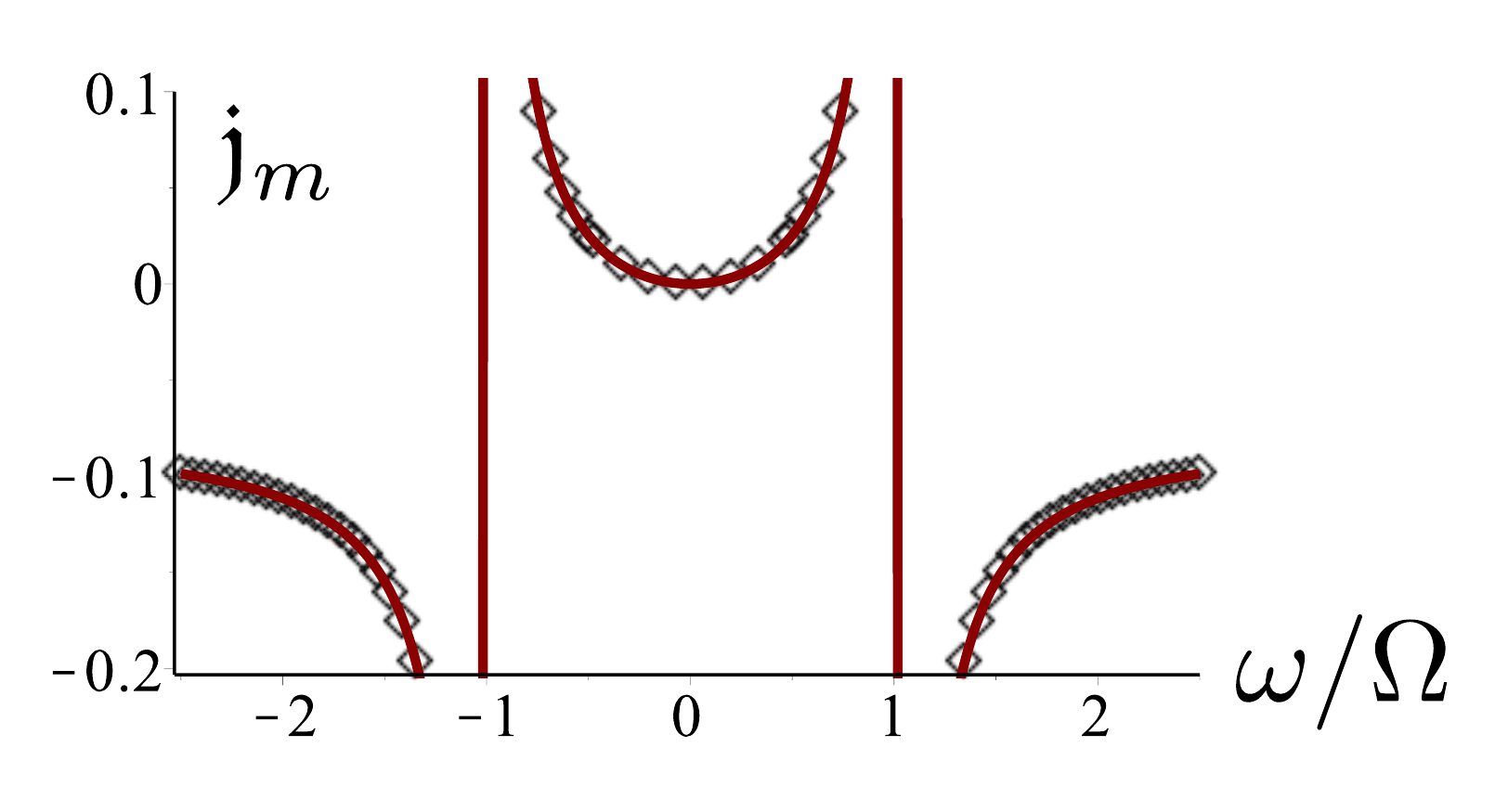}
\caption{Plot of $\gothj_m$ for $m=0$ as a function of $w$. The black diamonds result from the numerical evaluation of Eq.~(\ref{j_components}). The solid red curve is the mode-sum representation constructed in Sec.~\ref{sec:oscillator}. The vertical lines at $w \simeq \pm 1.0282$ correspond to the eigenfrequencies of dominant inertial modes (polar-led modes with $m=0$ and $\ell_0 = 2$, calculated for a $p = K\rho^2$ polytrope.)}  
\label{fig:j_m0} 
\end{figure} 

In Figs.~\ref{fig:j_m2}, \ref{fig:j_m1}, and \ref{fig:j_m0} we display plots of the dimensionless quadrupole moments $\gothj_m(\omega)$ for $m = 2$, $m = 1$, and $m=0$, respectively. The black diamonds correspond to a numerical evaluation of the integrals in Eqs.~(\ref{j_components}), calculated for the $p = K\rho^2$ polytropic model, for which $\hat{\rho} = (\pi/3) \sin(\pi \hat{r})/\hat{r}$. The solid red curves correspond to the construction detailed in Sec.~\ref{sec:oscillator}, based on a mode-sum representation of the tidal perturbation; the point is to demonstrate the excellent agreement between these different calculations. There are no black diamonds for $m=2$, because in this case the mode sum is exact and agrees precisely with Eq.~(\ref{j_components}). For $m=2$ the radial integration can be carried out analytically:
\begin{equation} 
\int_0^1 \hat{\rho}\, \hat{r}^6\, d\hat{r} = \frac{\pi^4 - 20\pi^2 + 120}{3\pi^4}. 
\label{rho-r6} 
\end{equation} 

The plots of $\gothj_m(\omega)$ feature simple poles when $\omega$ matches an eigenfrequency of a dominant inertial mode. For $m=2$ the pole occurs at $w = 4/3$, which corresponds to the eigenfrequency of an $r$-mode. For $m=1$ we have poles at $w \simeq -0.4130$ and $w \simeq 1.4014$; these are the eigenfrequencies of the polar-led inertial modes labelled $\ell_0 = 2$ by Lockitch and Friedman \cite{lockitch-friedman:99}. For $m=0$ the poles occur at $w \simeq \pm 1.0282$, the eigenfrequencies of the polar-led inertial modes also labelled $\ell_0 = 2$. A finer sampling of $\gothj_m(\omega)$ would reveal weaker poles associated with higher-order, subdominant inertial modes.  
 
\section{Mode-sum representation of the perturbation} 
\label{sec:oscillator} 

In this section we construct an alternative representation of the velocity perturbation in terms of a sum over normal modes. Our goals here are twofold. First, we demonstrate that a set of just four modes provides a very accurate approximation to the perturbation. Second, we introduce some of the essential ingredients required in a discussion of the dynamical impact of the resonances on a binary inspiral; this discussion will be taken up in Sec.~\ref{sec:dynamics}.  

\subsection{Formalism} 

Methods to represent a driven perturbation of a rotating fluid as a sum over normal modes are developed in Sec.~II and Appendix A of Ref.~\cite{schenk-etal:01}. We begin with a very brief summary of these techniques.  

The formalism makes central use of the Lagrangian displacement vector $\xi^a$, which is related to the velocity field $\delta v^a$ by Eq.~(\ref{xi_vs_dv}). It is useful to introduce a linear-algebra notation in which $\xi^a$ is mapped to an abstract vector $\xi$ in a Hilbert space, and the operation $\KK^a_{\ b} \xi^b:= v^b \nabla_b \xi^a$ is mapped to $\KK \xi$. We also introduce an inner product $\bkt{\eta}{\xi}$ between two vectors $\eta$ and $\xi$, defined by 
\begin{equation}
\bkt{\eta}{\xi} := \int_V \rho\, \bar{\eta}_a \xi^a\, dV, 
\end{equation}
where an overbar indicates complex conjugation. The inner product satisfies the usual property that $\bkt{\xi}{\eta}$ is the complex conjugate of $\bkt{\eta}{\xi}$. 

A normal mode is a solution to the perturbation equations (without an external force) of the specific form 
\begin{equation}
\xi^a(t,\bm{x}) = \zeta^a(\bm{x}) e^{-i\omega t}. 
\label{mode_def}
\end{equation} 
The perturbation equations become an eigenvalue problem for the frequency $\omega$ and mode functions $\zeta^a$. There is an infinity of solutions, and we label each solution with a mode index $K$; for simplicity we assume that the mode spectrum is discrete and nondegenerate, so that there is a unique mode $\xi_K$ for each eigenfrequency $\omega_K$.  

Next we consider the perturbation equations with an external force, represented by the abstract vector $f$. We wish to represent the solution $\xi$ in terms of a sum over modes. We adopt a ``phase-space'' representation and write 
\begin{equation} 
\xi  = \sum_K q_K(t)\, \zeta_K, \qquad  
\partial_t \xi = \sum_K (-i\omega_K) q_K(t)\, \zeta_K, 
\label{mode-sum} 
\end{equation} 
for some amplitudes $q_K(t)$. The perturbation equations imply that these are determined by 
\begin{equation} 
\dot{q}_K + i\omega_K q_K = i f_K := -\frac{\bkt{\zeta_K}{f}}{2i\omega_K N_K}, 
\label{q_eq} 
\end{equation} 
where 
\begin{equation} 
N_K := \bkt{\zeta_K}{\zeta_K} - \frac{1}{i\omega_K} \bkt{\zeta_K}{\KK \zeta_K} 
\label{Ndef} 
\end{equation} 
provides a notion of mode norm. (This is to be interpreted with caution: while $N_K$ is real, it is not positive-definite.) For $f = 0$ we would have that $q_K \propto \exp(-i\omega_K t)$, and the expression for $\partial_t \xi$ would follow from the one for $\xi$ by direct differentiation. The expressions, however, are independent when $f \neq 0$.  

The normal modes that are implicated in a gravitomagnetic tidal interaction are the Lockitch-Friedman inertial modes \cite{lockitch-friedman:99}, which are labelled by the azimuthal integer $m$ and an additional mode index $n$. Inertial modes with negative values of $m$ are related to those with positive values by 
\begin{equation} 
\omega_n^{-m} = -\omega_n^m, \qquad 
\zeta_n^{-m} = \bar{\zeta}_n^m. 
\label{m_negative} 
\end{equation} 
These properties allow us to fold the sum over negative values of $m$ into the positive values. To achieve this folding, we note that since $f$ is a real vector, $\bkt{\zeta_n^{-m}}{f}$ is the complex conjugate of $\bkt{\zeta_n^m}{f}$. We also note that $N_n^{-m} = N_n^{m}$ is a direct consequence of the definition of Eq.~(\ref{Ndef}). From all this it follows that $q_n^{-m} = \bar{q}_n^m$. We further take into account the fact that the $m=0$ modes always come in complex-conjugate pairs; one member has a positive frequency, the second member has a negative frequency of the same absolute value, and the mode functions are complex conjugates of each other. These observations imply that Eqs.~(\ref{mode-sum}) can be written as     
\begin{equation}
\xi = \sum_{m \geq 0} \psum_n \bigl[ q_n^m(t)\, \zeta_n^m + \mbox{cc} \bigr], \qquad
\partial_t \xi = \sum_{m \geq 0} \psum_n \bigl[ -i\omega_n^m q_n^m(t)\, \zeta_n^m + \mbox{cc} \bigr]; 
\label{mode_sum_inertial} 
\end{equation} 
the prime on the sum over $n$ reminds us that for $m = 0$, the sum is to include positive-frequency modes only. Inserting Eqs.~(\ref{mode_sum_inertial}) within Eq.~(\ref{xi_vs_dv}), we find that 
\begin{equation}
\delta v^a = \sum_{m \geq 0} \psum_n \bigl[ i(m\Omega - \omega_n^m) q_n^m(t)\, (\zeta_n^m)^a
+ \mbox{cc} \bigr] 
\label{dv_inertialmodesum} 
\end{equation}
is the required representation of the velocity perturbation in terms of a sum over inertial modes. 

In the same notation specialized to inertial modes, Eq.~(\ref{q_eq}) becomes
\begin{equation} 
\dot{q}_n^m + i\omega_n^m q_n^m = i f_n^m := -\frac{\bkt{\zeta_n^m}{f}}{2i\omega_n^m N_n^m}, 
\label{q_eq_inertial} 
\end{equation} 
and the mode norm is now
\begin{equation} 
N_n^m := \bkt{\zeta_n^m}{\zeta_n^m} - \frac{1}{i\omega_n^m} \bkt{\zeta_n^m}{\KK \zeta_n^m}.  
\label{Ndef_inertial} 
\end{equation} 

\subsection{Description of inertial modes} 
\label{sec:description} 

The decomposition of the external force $f$ given in Eq.~(\ref{fa_vharm}) implies that the only relevant modes are those with $m = \{0, 1, 2\}$. The corresponding mode functions are 
\begin{subequations} 
\label{mode_functions} 
\begin{align} 
(\zeta_n^2)^a &= R^2 \sum_\ell \biggl[ \frac{i}{r} (a_\ell^2)_n\, r^a Y_\ell^2 
+ i (b_\ell^2)_n\, (Y_\ell^2)^a + (c_\ell^2)_n (X_\ell^2)^a \biggr], \\
(\zeta_n^1)^a &= R^2 \sum_\ell \biggl[ \frac{1}{r} (a_\ell^1)_n\, r^a Y_\ell^1 
+ (b_\ell^1)_n\, (Y_\ell^1)^a - i (c_\ell^1)_n (X_\ell^1)^a \biggr], \\  
(\zeta_n^0)^a &= R^2 \sum_\ell \biggl[ \frac{i}{r} (a_\ell^0)_n\, r^a Y_\ell^0 
+ i (b_\ell^0)_n\, (Y_\ell^0)^a + (c_\ell^0)_n (X_\ell^0)^a \biggr],
\end{align} 
\end{subequations} 
where $(a_\ell^m)_n$, $(b_\ell^m)_n$, $(c_\ell^m)_n$ are functions of the radial coordinate. These are all real and dimensionless, and for $m > 0$ they satisfy 
\begin{equation} 
(a_\ell^{-m})_n = -(a_\ell^m)_n, \qquad 
(b_\ell^{-m})_n = -(b_\ell^m)_n, \qquad 
(c_\ell^{-m})_n = (c_\ell^m)_n
\end{equation} 
in order to enforce Eq.~(\ref{m_negative}). Factors of $i$ were inserted in Eqs.~(\ref{mode_functions}) to ensure that the overlap integrals $\bkt{\zeta_n^m}{f}$ are real. 

Differential equations for the radial functions are obtained by following the same strategy as in Sec.~\ref{sec:perturbation}. Here also we rely on a truncation of the system of equations, justified by the fact that for the dominant inertial modes, radial functions with the lowest value of $\ell$ are much larger than functions with higher values of $\ell$. More precisely stated, for the implicated modes we have that $|(a_5^m)_n| < |(a_3^m)_n| \ll |(a_1^m)_n|$, $|(b_5^m)_n| < |(b_3^m)_n| \ll |(b_1^m)_n|$, and $|(c_6^m)_n| < |(c_4^m)_n| \ll |(c_2^m)_n|$. The equations take the same form as those listed in Secs.~\ref{sec:eqns_m2}, \ref{sec:eqns_m1}, and \ref{sec:eqns_m0}, with all driving terms --- proportional to $(r/R)^3$ --- set to zero.   

One of the mode equations is identical to Eq.~(\ref{Aprime}), 
\begin{equation} 
r \frac{d(a_\ell^m)_n}{dr} + \biggl( 1 + \frac{r}{\rho} \frac{d\rho}{dr} \biggr) (a_\ell^m)_n 
- \ell(\ell+1) (b_\ell^m)_n = 0,    
\label{aprime} 
\end{equation} 
and it is easy to show that it implies 
\begin{equation} 
\rho r^3 \bigl[ 3 (a_\ell^m)_n + \ell(\ell+1) (b_\ell^m)_n \bigr] 
= \frac{d}{dr} \bigl[ \rho r^4 (a_\ell^m)_n \bigr]. 
\end{equation} 
Integrating both sides from $r=0$ to $r=R$, we arrive at the identity 
\begin{equation} 
\int_0^R \rho r^3 \bigl[ 3 (a_\ell^m)_n + \ell(\ell+1) (b_\ell^m)_n \bigr]\, dr = 0, 
\label{ab_identity} 
\end{equation} 
which will be required below. On the right-hand side, the boundary term at $r=0$ vanishes because of the factor of $r^4$, and the one at $r=R$ vanishes because $\rho = 0$ for our stellar models, and also because $(a_\ell^m)_n = 0$, by virtue of the mode version of Eq.~(\ref{BC_A}). 

Numerical solutions to the eigenvalue problem for the mode frequencies $\omega_n^m$ and mode functions $(\zeta_n^m)^a$ are obtained with the computational methods described in Sec.~\ref{sec:solution}. We develop series expansions near $r=0$ and $r=R$, and we integrate outward from $r \simeq 0$ and inward from $r \simeq R$. To determine the eigenfrequencies and the free coefficients that appear in the series expansions, we search for solutions that match smoothly at a middle point $r^\sharp$. Our results are in excellent agreement with those of Lockitch and Friedman \cite{lockitch-friedman:99}.  

\subsection{Overlap integrals} 
\label{sec:overlap} 

We insert Eqs.~(\ref{fa_vharm}) and (\ref{mode_functions}) within the overlap integrals $\bkt{\zeta_n^m}{f}$, and perform  the angular integration with the help of the orthogonality relations of Eq.~(\ref{ortho1}) and (\ref{ortho2}). We find 
\begin{align} 
\bkt{\zeta_n^m}{f} &=\alpha^m \frac{{\cal B}_0}{c^2} R^2 \sum_\ell \int \rho \biggl\{ 
(a_\ell^m)_n \Bigl[ (f_\ell^m)^{\rm R} e^{-i\omega t} + (-1)^m (\bar{f}_\ell^{-m})^{\rm R} e^{i\omega t} \Bigr] 
\nonumber \\ & \quad \mbox{} 
+ \ell(\ell+1) (b_\ell^m)_n \Bigl[ (f_\ell^m)^{\rm P} e^{-i\omega t} 
+ (-1)^m (\bar{f}_\ell^{-m})^{\rm P} e^{i\omega t} \Bigr] 
\nonumber \\ & \quad \mbox{} 
+ i\ell(\ell+1) (c_\ell^m)_n \Bigl[ (f_\ell^m)^{\rm A} e^{-i\omega t} 
+ (-1)^m (\bar{f}_\ell^{-m})^{\rm A} e^{i\omega t} \Bigr] \biggr\}\, dr, 
\end{align} 
where $\alpha^m = -i$ when $m=0, 2$, and $\alpha^m = 1$ when $m=1$. Next we exploit the observation of Eq.~(\ref{RP-proportionality}) and the identity of Eq.~(\ref{ab_identity}) to eliminate the integrals involving the radial (R) and polar (P) components of the external force. And because the axial (A) components come with $\ell = 2$ only, the overlap integrals reduce to 
\begin{equation} 
\bkt{\zeta_n^m}{f} = 6i\alpha^m \frac{{\cal B}_0}{c^2} R^2 \int \rho (c_2^m)_n 
\Bigl[ (f_2^m)^{\rm A} e^{-i\omega t} 
+ (-1)^m (\bar{f}_2^{-m})^{\rm A} e^{i\omega t} \Bigr]\, dr. 
\end{equation} 
Making the substitutions from Eq.~(\ref{fa_axial}), we arrive at 
\begin{subequations} 
\label{overlap1} 
\begin{align} 
\bkt{\zeta_n^2}{f} &= \frac{{\cal B}_0}{c^2} M R^3\, \gothp_n^2 \bigl[ 
-\sin\iota(\cos\iota+1)(2\Omega-\omega) e^{-i\omega t} 
+ \sin\iota(\cos\iota-1)(2\Omega+\omega) e^{i\omega t} \bigr], \\
\bkt{\zeta_n^1}{f} &= \frac{{\cal B}_0}{c^2} M R^3\, \gothp_n^1 \bigl[ 
(\cos\iota+1)(2\cos\iota-1)(\Omega-\omega) e^{-i\omega t} 
- (\cos\iota-1)(2\cos\iota+1)(\Omega+\omega) e^{i\omega t} \bigr], \\ 
\bkt{\zeta_n^0}{f} &= \frac{{\cal B}_0}{c^2} M R^3\, \gothp_n^0\, \sin\iota\cos\iota\, \omega 
(e^{-i\omega t} + e^{i\omega t}) 
\end{align} 
\end{subequations} 
with 
\begin{subequations} 
\label{overlap2} 
\begin{align} 
\gothp_n^2 &:= 3 \sqrt{\frac{3}{10\pi}} \int \hat{\rho}\, (c_2^2)_n\, \hat{r}^3\, d\hat{r}, \\ 
\gothp_n^1 &:= 3 \sqrt{\frac{3}{10\pi}} \int \hat{\rho}\, (c_2^1)_n\, \hat{r}^3\, d\hat{r}, \\ 
\gothp_n^0 &:= \frac{9}{\sqrt{5\pi}} \int \hat{\rho}\, (c_2^0)_n\, \hat{r}^3\, d\hat{r}, 
\label{gothp2_def} 
\end{align}
\end{subequations} 
where $\hat{\rho} := (4\pi R^3/3M) \rho$ and $\hat{r} := r/R$. The reduced overlap integrals $\gothp_n^m$ are dimensionless.  

\subsection{Implicated modes} 
\label{sec:implicated} 

Equations (\ref{overlap2}) indicate that the inertial modes implicated in the overlap integrals are those for which $(c_2^m)_n \neq 0$. Adopting the terminology of Lockitch and Friedman \cite{lockitch-friedman:99}, this restriction selects the following set of modes: 
\begin{itemize} 
\item For $m = 2$ we have axial-led modes with nonvanishing $(a_3^2)_n$, $(a_5^2)_n$, $(a_7^3)_n$ and so on, $(b_3^2)_n$, $(b_5^2)_n$, $(b_7^2)_n$ and so on, and $(c_2^2)_n$, $(c_4^2)_n$, $(c_6^2)_n$ and so on. But as we are about to show, there is only one mode that produces a nonvanishing overlap integral with the external force: the $r$-mode with frequency $\omega_\bullet^2 = (4/3)\Omega$ and sole nonvanishing radial function $(c_2^2)_\bullet = (r/R)^3$. For a barotropic star, the $r$-mode is independent of the equation of state.   
\item For $m = 1$ we have polar-led modes with nonvanishing $(a_1^1)_n$, $(a_3^1)_n$, $(a_5^1)_n$ and so on, $(b_1^1)_n$, $(b_3^1)_n$, $(b_5^1)_n$ and so on, and $(c_2^1)_n$, $(c_4^1)_n$, $(c_6^1)_n$ and so on. The two dominant modes are those designated by $\ell_0 = 2$, and for the polytropic model $p = K\rho^2$, they have the eigenfrequencies $\omega_{\rm I}^1 \simeq 1.4014\, \Omega$ and  $\omega_{\rm II}^1 \simeq -0.4130\, \Omega$.  
\item For $m = 0$ we have polar-led modes with nonvanishing $(a_1^0)_n$, $(a_3^0)_n$, $(a_5^0)_n$ and so on, $(b_1^0)_n$, $(b_3^0)_n$, $(b_5^0)_n$ and so on, and $(c_2^0)_n$, $(c_4^0)_n$, $(c_6^0)_n$ and so on. The two dominant modes are also those with $\ell_0 = 2$, and for the polytropic model they have frequencies $\omega_{\rm I}^0 \simeq 1.0282\, \Omega$ and $\omega_{\rm II}^0 \simeq -1.0282\, \Omega$. As was discussed previously, only the positive-frequency mode is included in the mode sum. 
\end{itemize} 
The dominant modes for $m=1$ and $m=0$ are stable; the $r$-mode for $m=2$ is subjected to the Chandrasekhar-Friedman-Schutz instability \cite{chandrasekhar:70, friedman-schutz:78b, andersson:98, friedman-morsink:98}. 

We shall now establish that  
\begin{equation} 
\int \rho\, (c_2^2)_n\, r^3\, dr = 0 
\label{rmode_only} 
\end{equation} 
for all inertial modes except for the $r$-mode. As was stated previously, this implies that for $m=2$, only the $r$-mode is implicated in the decomposition of the velocity perturbation. The property follows from the orthogonality of normal modes with distinct frequencies. 

The statement of orthogonality for modes $\zeta_K$ and $\zeta_{K'}$ with $\omega_K \neq \omega_{K'}$ is [Eq.~(2.32) of Ref.~\cite{schenk-etal:01}]
\begin{equation} 
(\omega_K + \omega_{K'}) \bkt{\zeta_K}{\zeta_{K'}} + 2i \bkt{\zeta_K}{\KK \zeta_{K'}} = 0. 
\label{ortho_KK} 
\end{equation}  
In this we insert for $\zeta_K$ a generic inertial mode described by 
\begin{equation} 
\zeta_K^a = R^2 \sum_\ell \biggl[ \frac{i}{r} (a_\ell^2)_n\, r^a Y_\ell^2 
+ i (b_\ell^2)_n\, (Y_\ell^2)^a + (c_\ell^2)_n\, (X_\ell^2)^a \biggr],  
\end{equation} 
and for $\zeta_{K'}$ the $r$-mode with eigenfrequency $\omega_{K'} = (4/3)\Omega$ and mode function $\zeta^a_{K'} = R^2 (r/R)^3 (X_2^2)^a$. For this mode we have that 
\begin{equation} 
(\KK \zeta_{K'})^a = v^b \nabla_b \zeta_{K'}^b = \Omega R^2 (r/R)^3 \biggl[ \frac{2\sqrt{7}}{7} \frac{1}{r}\, r^a Y_3^2 
+ \frac{2\sqrt{7}}{21} (Y_3^2)^a + \frac{5i}{3} (X_2^2)^a \biggr]. 
\end{equation} 
Using the orthogonality relations of Eq.~(\ref{ortho1}) and (\ref{ortho2}), we find that 
\begin{equation} 
\bkt{\zeta_K}{\zeta_{K'}} = 6R \int \rho\, (c_2^2)_n\, r^3\, dr
\end{equation} 
and 
\begin{equation} 
\bkt{\zeta_K}{\KK \zeta_{K'}} = 
-\frac{2i\sqrt{7}}{21} \Omega R \int \rho r^3 \bigl[ 3(a_3^2)_n + 12(b_3^2)_n \bigr]\, dr 
+ 10 i \Omega R \int \rho\, (c_2^2)_n\, r^3\, dr. 
\end{equation} 
This simplifies to 
\begin{equation} 
\bkt{\zeta_K}{\KK \zeta_{K'}} = 10 i \Omega R \int \rho\, (c_2^2)_n\, r^3\, dr 
\end{equation} 
after invoking Eq.~(\ref{ab_identity}). The statement of orthogonality therefore becomes 
\begin{equation} 
(\omega_K - 2\Omega) \int \rho\, (c_2^2)_n\, r^3\, dr = 0, 
\end{equation} 
and we arrive at Eq.~(\ref{rmode_only}) whenever $\omega_K \neq 2\Omega$. 

With Eq.~(\ref{rmode_only}) established, we conclude that for $m = 2$, the sums over $n$ in Eqs.~(\ref{mode_sum_inertial}) reduce to a single term, which we denote $q_\bullet^2\, \zeta_\bullet^2 + \mbox{cc}$ and $-i\omega_\bullet^2\, q_\bullet^2\, \zeta_\bullet^2 + \mbox{cc}$, respectively. Equation (\ref{gothp2_def}) becomes  
\begin{equation} 
\gothp_\bullet^2 = 3 \sqrt{\frac{3}{10\pi}} \int \hat{\rho}\, \hat{r}^6\, d\hat{r};
\label{gothp2_result} 
\end{equation} 
the integral was evaluated in Eq.~(\ref{rho-r6}) for the $p = K\rho^2$ polytropic model. 

\subsection{Mode norm} 
\label{sec:norm} 

The mode norm $N_n^m$ is defined by Eq.~(\ref{Ndef_inertial}), and to calculate it we introduce the notation 
\begin{equation} 
\bkt{\zeta_n^m}{\zeta_n^m} = \frac{3}{4\pi} M R^2\, \gothn_n^m, \qquad 
\bkt{\zeta_n^m}{\KK\zeta_n^m} = \frac{3i}{4\pi} M R^2\Omega\, \gothm_n^m,
\end{equation} 
in which $\gothn_m^n$ and $\gothn_m^n$ are dimensionless. With this we have that 
\begin{equation} 
N_n^m = \frac{3}{4\pi} M R^2\, \hat{N}_n^m,  
\end{equation} 
where 
\begin{equation} 
\hat{N}_n^m := \gothn_n^m - \frac{\gothm_n^m}{w_n^m}, \qquad 
w_n^m := \omega_n^m/\Omega. 
\label{Nhat_def} 
\end{equation} 

For $m=2$ we have the exact expressions 
\begin{equation} 
\gothn_\bullet^2 = 6 \int_0^1 \hat{\rho}\, \hat{r}^6\, d\hat{r}, \qquad 
\gothm_\bullet^2 = 10 \int_0^1 \hat{\rho}\, \hat{r}^6\, d\hat{r}, \qquad 
\hat{N}_\bullet^2 = -\frac{3}{2} \int_0^1 \hat{\rho}\, \hat{r}^6\, d\hat{r}. 
\label{gothnm_m2} 
\end{equation}  
To calculate $\gothn_n^m$ and $\gothm_n^m$ for $m=0$ and $m=1$ we truncate the description of the mode to include only the dominant functions $a_1^m$, $b_1^m$, and $c_2^m$. This gives 
\begin{subequations}
\label{gothnm_m1} 
\begin{align} 
\gothn_n^1 &= \int \hat{\rho} \bigl[ (a_1^1)_n^2 + 2 (b_1^1)_n^2 + 6 (c_2^1)_n^2 \bigr]\, d\hat{r}, \\ 
\gothm_n^1 &= \int \hat{\rho} \Bigl\{ (a_1^1)_n^2 - 2 (a_1^1)_n (b_1^1)_n + (b_1^1)_n^2 
+ \frac{6\sqrt{5}}{5} \bigl[ (a_1^1)_n  - (b_1^1)_n \bigr] (c_2^1)_n + 5(c_2^1)_n^2 \Bigr\}\,  d\hat{r} 
\end{align} 
\end{subequations} 
and 
\begin{subequations}
\label{gothnm_m0} 
\begin{align} 
\gothn_n^0 &= \int \hat{\rho} \bigl[ (a_1^0)_n^2 + 2 (b_1^0)_n^2 + 6 (c_2^0)_n^2 \bigr]\, d\hat{r}, \\ 
\gothm_n^0 &= \frac{4\sqrt{15}}{5} \int \hat{\rho} \bigl[ (a_1^0)_n - (b_1^0)_n \bigr] (c_2^0)_n\, d\hat{r}. 
\end{align} 
\end{subequations} 

\subsection{Stellar models}

We compute eigenfrequencies $w_n^m := \omega_n^m/\Omega$, overlap integrals $\gothp_n^m$, and mode norms $\gothn_n^m$, $\gothm_n^m$, and $\hat{N}_n^m$ for a variety of stellar models. These are defined by the density function 
\begin{equation}
\rho = C \biggl[ \frac{\sin(\pi r/R)}{(\pi r/R)} \biggr]^k,
\label{rho_kmodel} 
\end{equation}
in which $C$ is a constant (irrelevant for our purposes) and $k$ is an integer in the set $\{ 0, 1, 2, 3 \}$. With $k=0$ we have a star of constant density, and $k = 1$ corresponds to a polytrope with $p = K\rho^2$. To the remaining values of $k$ we can assign an equation of state by obtaining $p(r)$ from the structure equations and expressing the result as $p(r(\rho))$, where $r(\rho)$ is the inversion of Eq.~(\ref{rho_kmodel}). The stellar models described by Eq.~(\ref{rho_kmodel}) range from being nearly uniform (for small values of $k$) to being centrally dense (for large values of $k$). We recall that $\rho$ appears explicitly in Eqs.~(\ref{overlap2}), (\ref{gothnm_m1}), and (\ref{gothnm_m0}), but that it also enters --- via the function $1 + (r/\rho) (d\rho/dr)$ --- in the mode equations.   

\begin{table} 
\caption{\label{tab:overlaps} Eigenfrequencies, overlap integrals, and norms for the relevant modes, calculated for the stellar models described by Eq.~(\ref{rho_kmodel}).}  
\begin{ruledtabular} 
\begin{tabular}{cccccccc}
$k$ & $m$ & mode & $w_n^m$ & $\gothp_n^m$ & $\gothn_n^m$ & $\gothm_n^m$ & $\hat{N}_n^m$ \\ 
\hline
  0 & 2 & $\bullet$ & 1.3333 & 1.3244e-1 & 8.5714e-1 & 1.4286 & -2.1429e-1 \\
    & 1 & I & 1.1766 & 1.3244e-1 & 1.2286 & 1.3371 & 9.2208e-2 \\
    & 1 & II & -5.0994e-1 & 1.3244e-1 & 2.8349 & 6.9463e-1 & 4.1970 \\
    & 0 & I &  8.9443e-1 & 3.2440e-1 & 1.7143 & 7.6665e-1 & 8.5714e-1 \\
\hline
  1 & 2 & $\bullet$ & 1.3333 & 6.3502e-2 & 4.1099e-1 & 6.8498e-1 & -1.0275e-1 \\
    & 1 & I & 1.4014 & -4.1974e-2 & 3.0282e-1 & 3.6375e-1 & 4.3258e-2 \\
    & 1 & II & -4.1300e-1 & 2.6417e-2 & 1.7609e-1 & 5.1716e-2 & 3.0131e-1 \\
    & 0 & I & 1.0282 & -7.9947e-2 & 2.1767e-1 & 1.1196e-1 & 1.0879e-1 \\
\hline
  2 & 2 & $\bullet$ & 1.3333 & 3.5252e-2 & 2.2816e-1 & 3.8026e-1 & -5.7039e-2 \\
    & 1 & I & 1.4979 & -3.1348e-2 & 3.2385e-1 & 4.0454e-1 & 5.3785e-2 \\
    & 1 & II & -3.8059e-1 & 2.2777e-2 & 2.1528e-1 & 6.6684e-2 & 3.9049e-1 \\
    & 0 & I & 1.0763 & -6.3472e-2 & 2.4632e-1 & 1.3269e-1 & 1.2303e-1 \\
\hline
  3 & 2 & $\bullet$ & 1.3333 & 2.1933e-2 & 1.4195e-1 & 2.3659e-1 & -3.5489e-2 \\
    & 1 & I & 1.5466 & -2.4483e-2 & 3.2772e-1 & 4.1736e-1 & 5.7869e-2 \\
    & 1 & II & -3.6583e-1 & 1.9210e-2 & 2.3666e-1 & 7.5054e-2 & 4.4182e-1 \\
    & 0 & I & 1.1003 & -5.2790e-2 & 2.7400e-1 & 1.5080e-1 & 1.1369e-1
\end{tabular} 
\end{ruledtabular} 
\end{table} 

Our numerical results are displayed in Table~\ref{tab:overlaps}. It is interesting to see how the mode frequency varies with the parameter $k$: while the $r$-mode frequency is a universal $w^2_\bullet = 4/3$, all other frequencies increase with $k$, that is, increase as the star becomes centrally dense. 

The entries for $k = 0$ in Table~\ref{tab:overlaps} can be obtained analytically, using results first presented by Lockitch and Friedman \cite{lockitch-friedman:99}. For $m=1$ the mode frequencies are solutions to $15w^2 - 10w - 9 = 0$; we have  $w^1_{\rm I} = 1/3 + 4\sqrt{10}/15 \simeq 1.1766$ and $w^1_{\rm II} = 1/3 - 4\sqrt{10}/15 \simeq -0.5099$, and the corresponding mode functions are ($n = \{ \mbox{I}, \mbox{II} \}$) 
\begin{equation}
(a_1^1)_n = -\frac{\sqrt{5}}{3} (3 w_n^1 - 2) \frac{r(r^2-R^2)}{R^3}, \qquad
(b_1^1)_n = -\frac{\sqrt{5}}{3} (3 w_n^1 - 2) \frac{r(2r^2-R^2)}{R^3}, \qquad
(c_2^1)_n = (r/R)^3.
\end{equation}
For $m=0$ the mode frequencies are solutions to $5w^2 - 4 = 0$, and the positive solution is $w_{\rm I}^0 = 2/\sqrt{5} \simeq 0.8944$. The corresponding mode functions are ($n = \{ \mbox{I}, \mbox{II} \}$, with $n = \mbox{II}$ designating the negative-frequency mode) 
\begin{equation}
(a_1^0)_n = -\frac{\sqrt{15}}{2} w_n^0 \frac{r(r^2-R^2)}{R^3}, \qquad
(b_1^0)_n = -\frac{\sqrt{15}}{2} w_n^0 \frac{r(2r^2-R^2)}{R^3}, \qquad
(c_2^0)_n = (r/R)^3.
\end{equation}

The entries that pertain to the $r$-mode can also be calculated analytically, for all the selected values of $k$. For all other modes, and for $k \neq 0$, the eigenfrequencies and mode functions must be obtained numerically, as was described in Sec.~\ref{sec:description}. The results for $\gothp_n^m$, $\gothn_n^m$, and $\gothm_n^m$ also require a numerical computation, and their values reflect a choice of normalization for the mode. We chose to normalize them with the condition $(b_1^m)_n(r=R) = \frac{1}{2}(1+k)$.  

\subsection{Amplitudes, velocity field, and current quadrupole moment} 

Collecting the foregoing results, we have that the forcing functions $f_n^m$ defined by Eq.~(\ref{q_eq_inertial}) can be expressed as
\begin{equation}
f_n^m = A^m_{+n} e^{-i\omega t} + A^m_{-n} e^{i\omega t}
\label{f_nm_fixedorbit}
\end{equation}
with
\begin{subequations}
\label{Apm}
\begin{align}
A^2_{\pm \bullet} &= \mp \frac{{\cal B}_0 R}{c^2} \frac{2\pi \gothp^2_\bullet}{3 \hat{N}^2_\bullet \omega^2_\bullet}
\sin\iota (\cos\iota \pm 1) (2\Omega \mp \omega), \\
A^1_{\pm n} &= \pm \frac{{\cal B}_0 R}{c^2} \frac{2\pi \gothp^1_n}{3 \hat{N}^1_n \omega^1_n}
(\cos\iota \pm 1)(2\cos\iota \mp 1) (\Omega \mp \omega), \\
A^0_{\pm n} &= \frac{{\cal B}_0 R}{c^2} \frac{2\pi \gothp^0_n}{3 \hat{N}^0_n \omega^0_n}
\sin\iota \cos\iota\, \omega. 
\end{align}
\end{subequations} 
The solution to Eq.~(\ref{q_eq_inertial}) is then
\begin{equation}
q_n^m = \frac{A_{+n}^m}{\omega_n^m - \omega} e^{-i\omega t}
+ \frac{A_{-n}^m}{\omega_n^m + \omega} e^{i\omega t}.
\label{oscillators_fixed} 
\end{equation}

We insert Eq.~(\ref{oscillators_fixed}) within Eq.~(\ref{dv_inertialmodesum}) for the velocity perturbation, and obtain 
\begin{equation} 
\delta v^a = (\delta v^a)_{m=0} + (\delta v^a)_{m=1} + (\delta v^a)_{m=2}
\label{dv_modesum} 
\end{equation} 
with 
\begin{subequations} 
\label{dv_m} 
\begin{align} 
(\delta v^a)_{m=2} &= i\frac{{\cal B}_0}{c^2} R^3 
\frac{2\pi \gothp_\bullet^2(2-w_\bullet^2)}{3 w_\bullet^2 \hat{N}_\bullet^2} 
\biggl[ -\sin\iota (\cos\iota+1) \frac{2-w}{w_\bullet^2 - w} e^{-i\omega t} 
+ \sin\iota (\cos\iota-1) \frac{2+w}{w_\bullet^2 + w} e^{i\omega t} \biggr] 
\nonumber \\ & \quad \mbox{} 
\times (r/R)^3\, (X_2^2)^a + \mbox{cc}, \\ 
 (\delta v^a)_{m=1} &= \frac{{\cal B}_0}{c^2} R^3 \sum_{n= {\rm I}, {\rm II}} 
\frac{2\pi \gothp_n^1 (1-w_n^1)}{3 w_n^1 \hat{N}_n^1} 
\biggl[ (\cos\iota+1)(2\cos\iota-1) \frac{1-w}{w_n^1 - w} e^{-i\omega t} 
- (\cos\iota-1)(2\cos\iota+1) \frac{1+w}{w_n^1 + w} e^{i\omega t} \biggr] 
\nonumber \\ & \quad \mbox{} \times 
\sum_\ell \biggl[ \frac{i}{r} (a_\ell^1)_n\, r^a Y_\ell^1 
+ i (b_\ell^1)_n\, (Y_\ell^1)^a + (c_\ell^1)_n\, (X_\ell^1)^a \biggr] + \mbox{cc}, \\ 
(\delta v^a)_{m=0} &= \frac{{\cal B}_0}{c^2} R^3 \frac{2\pi \gothp_{\rm I}^0}{3 \hat{N}_{\rm I}^0} 
\biggl[ \sin\iota\cos\iota \frac{w}{w_{\rm I}^0 - w} e^{-i\omega t} 
+ \sin\iota\cos\iota \frac{w}{w_{\rm I}^0 + w} e^{i\omega t} \biggr] 
\nonumber \\ & \quad \mbox{} \times 
\sum_\ell \biggl[ \frac{1}{r} (a_\ell^0)_{\rm I}\, r^a Y_\ell^0 
+ (b_\ell^0)_{\rm I}\, (Y_\ell^0)^a - i (c_\ell^0)_{\rm I}\, (X_\ell^0)^a \biggr] + \mbox{cc},
\end{align} 
\end{subequations} 
where $w := \omega/\Omega$ and $w_n^m := \omega_n^m/\Omega$.  

Next we involve the velocity field of Eq.~(\ref{dv_m}) in a calculation of the current quadrupole moment $J^{jk}$, just as we did back in Sec.~\ref{sec:quadmoment}. First we return to Eq.~(\ref{dv_inertialmodesum}), which we insert within Eq.~(\ref{J_def}). We make use of the identities of Eqs.~(\ref{iden}), and observe that the surviving integrals are those displayed in Eq.~(\ref{overlap2}). We arrive at
\begin{equation}
J^{jk} = \sum_{m\geq 0} \psum_n (J^{jk})_n^m
\label{Jjk_mode1} 
\end{equation}
with
\begin{equation}
(J^{jk})_n^m = \frac{6i}{5} \frac{\alpha^m}{\beta^m} M R^3 (m\Omega - \omega_n^m) \gothp_n^m q_n^m 
(\bar{\scrpt Y}_2^m)^{jk} + \mbox{cc}, 
\label{Jjk_mode2}
\end{equation} 
where $\alpha^m = 1$ when $m=0, 2$ and $\alpha^m = -i$ when $m = 1$, and where $\beta^m$ denotes the numerical coefficients in front of the integrals in Eq.~(\ref{overlap2}). More explicitly, we have that
\begin{subequations}
\label{Jjk_mode3}
\begin{align}
(J^{jk})_\bullet^2 &= \frac{1}{2} M R^3 (2\Omega - \omega^2_\bullet) \gothp^2_\bullet \left[
q_\bullet^2 \left( 
\begin{array}{ccc} 
i  & -1 & 0 \\
-1 & -i & 0 \\ 
0 & 0 & 0 
\end{array} \right)
+ \mbox{cc} \right], \\
(J^{jk})_n^1 &= \frac{1}{2} M R^3 (\Omega - \omega^1_n) \gothp^1_n \left[
q_n^1 \left( 
\begin{array}{ccc} 
0  & 0 & -1 \\
0 & 0 & -i \\ 
-1 & -i & 0 
\end{array} \right)
+ \mbox{cc} \right], \\
(J^{jk})_n^0 &= \frac{1}{6} M R^3 (-\omega^0_n) \gothp^0_n \left[
q_n^0 \left( 
\begin{array}{ccc} 
-i  & 0 & 0 \\
0 & -i & 0 \\ 
0 & 0 & 2i 
\end{array} \right)
+ \mbox{cc} \right].
\end{align}
\end{subequations} 
If we now insert Eq.~(\ref{oscillators_fixed}) with Eqs.~(\ref{Apm}) and re-express Eq.~(\ref{Jjk_mode1}) in the form of 
Eq.~(\ref{J_decomposed}), we arrive at Eq.~(\ref{J_components}) with 
\begin{subequations} 
\label{j_modesum} 
\begin{align} 
\gothj_{\pm 2}(\omega) &= -\frac{2\pi}{3} 
\frac{(\gothp_\bullet^2)^2}{\hat{N}_\bullet^2} \frac{2-w_\bullet^2}{w_\bullet^2} 
\frac{2 \mp w}{w_\bullet^2 \mp w}, 
\label{gothj_modesum_m2} \\
\gothj_{\pm 1}(\omega) &= \pm \sum_{n = {\rm I}, {\rm II}} \frac{2\pi}{3} 
\frac{(\gothp_n^1)^2}{\hat{N}_n^1} \frac{1-w_n^1}{w_n^1} 
\frac{1 \mp w}{w_n^1 \mp w}, \\  
\gothj_{0}(\omega) &= \frac{4\pi}{9} 
\frac{(\gothp_{\rm I}^0)^2}{\hat{N}_{\rm I}^0} 
\frac{w^2}{(w_{\rm I}^0-w) (w_{\rm I}^0+w)}. 
\end{align} 
\end{subequations} 
These alternative expressions for $\gothj_1(\omega)$ and $\gothj_0(\omega)$ are plotted in Figs.~\ref{fig:j_m1} and \ref{fig:j_m0} for $k=1$ (the polytrope with $p = K\rho^2$), along with the results obtained in Sec.~\ref{sec:quadmoment}. The previous results are shown as black diamonds, while the mode-sum representations are plotted as solid red lines. We see that the numerical agreement is excellent, and conclude that even with just a handful of dominant modes, the mode-sum representation of the velocity perturbation is very accurate. For $\gothj_{\pm 2}(\omega)$ the agreement is actually exact: with $w_\bullet^2 = 4/3$, $\gothp_\bullet^2$ given by Eq.~(\ref{gothp2_result}), and $\hat{N}_\bullet^2$ by Eq.~(\ref{gothnm_m2}), we find that Eq.~(\ref{gothj_modesum_m2}) becomes precisely identical to Eq.~(\ref{j_components_m2}).  

\section{Dynamical impact of gravitomagnetic tidal resonances} 
\label{sec:dynamics} 

In this section we calculate the impact on a binary inspiral of the gravitomagnetic tidal resonances identified in Sec.~\ref{sec:solution}. Our end goal is to obtain the accumulated gravitational-wave phase shift produced by the resonances. We rely heavily on the methods developed by Flanagan and Racine \cite{flanagan-racine:07} in their exploration of the $r$-mode resonance.  

\subsection{Inspiral}
\label{subsec:inspiral} 

We consider a binary system on a circular orbit, driven into an inspiral by gravitational radiation reaction. We have that the mass of the reference star is $M$, the mass of the companion is $M'$, the orbital angular velocity is $\omega$, and the orbital radius is $p =(GM_{\rm tot})^{1/3} \omega^{-2/3}$, where $M_{\rm tot} = M + M'$ is the total mass. At leading order in a post-Newtonian description of the inspiral, the rate of change of the orbital radius is given by [Eq.~(12.251) of Ref.~\cite{poisson-will:14}]
\begin{equation}
\dot{p} = -\frac{64}{5} c^{-5} (G {\cal M})^{5/3} (G M_{\rm tot})^{4/3} p^{-3},
\label{pdot}
\end{equation}
where an overdot indicates differentiation with respect to time, and
\begin{equation} 
{\cal M} := \frac{ (M M')^{3/5} }{(M + M')^{1/5} }
\label{chirp_mass}
\end{equation}
is the binary's chirp mass. The corresponding rate of change of the angular velocity is 
\begin{equation}
\dot{\omega} = \frac{96}{5} c^{-5} (G {\cal M})^{5/3} \omega^{11/3}. 
\label{omegadot}
\end{equation}

We introduce a reference frequency $\omega_0$ (later to be identified with one of the resonant frequencies) and a radiation-reaction time scale defined by $t_{\rm rr} := (\omega/\dot{\omega})_{\omega=\omega_0}$. This is 
\begin{equation}
t_{\rm rr} = \frac{5}{96} c^5 (G {\cal M})^{-5/3} \omega_0^{-8/3}
= 9.1 \times 10^{-1} \biggl( \frac{1.2\ M_\odot}{\cal M} \biggr)^{5/3}
\biggl( \frac{100\ \mbox{Hz}}{\omega_0/2\pi} \biggr)^{8/3}\ \mbox{s};
\label{trr_def} 
\end{equation}
in the second expression we inserted fiducial values for the binary parameters; a chirp mass of $1.2\ M_\odot$ corresponds to an equal-mass binary with $M = M' = 1.4\ M_\odot$.

The solutions to Eqs.~(\ref{pdot}) and (\ref{omegadot}) are
\begin{equation}
p(t) = p_0 (1 - \tfrac{8}{3} \tau)^{1/4}, \qquad
\omega(t) = \omega_0 (1 - \tfrac{8}{3} \tau)^{-3/8},
\label{p_omega}
\end{equation}
where $\tau := (t-t_0)/t_{\rm rr}$ and $p_0 := (GM_{\rm tot})^{1/3} \omega_0^{-2/3}$; $t_0$ is the time at which $p = p_0$ and $\omega = \omega_0$. The orbital phase $\Phi$ is the integral of $\omega$ with respect to time, and this is given by
\begin{equation}
\Phi(t) = \Phi_0 + \frac{3}{5} \omega_0 t_{\rm rr} \bigl[ 1 - (1 - \tfrac{8}{3} \tau)^{5/8} \bigr],
\label{orb_phase}
\end{equation}
where $\Phi_0$ is the phase at time $t = t_0$.

For future reference we introduce a resonance-transit time scale $t_{\rm res} := (\pi t_{\rm rr}/\omega_0)^{1/2}$, the geometrical mean between the orbital period at resonance and the radiation-reaction time. This is 
\begin{equation}
t_{\rm res} = \sqrt{\frac{5\pi}{96}} c^{5/2} (G {\cal M})^{-5/6} \omega_0^{-11/6}
= 6.7 \times 10^{-2} \biggl( \frac{1.2\ M_\odot}{\cal M} \biggr)^{5/6}
\biggl( \frac{100\ \mbox{Hz}}{\omega_0/2\pi} \biggr)^{11/6}\ \mbox{s}. 
\label{tres_def}
\end{equation}
As the name indicates, $t_{\rm res}$ provides a measure of the time required for the inspiral to transit through a resonance at frequency $\omega_0$. The ratio of time scales is $\epsilon := t_{\rm res}/t_{\rm rr}$, or
\begin{equation}
\epsilon = \sqrt{\frac{96\pi}{5}} c^{-5/2} (G {\cal M})^{5/6} \omega_0^{5/6}
= 7.4 \times 10^{-2} \biggl( \frac{\cal M}{1.2\ M_\odot} \biggr)^{5/6}
\biggl( \frac{\omega_0/2\pi}{100\ \mbox{Hz}} \biggr)^{5/6}. 
\label{epsilon_def}
\end{equation}
We see that $\epsilon \ll 1$ for binaries of interest, and we shall rely on this property in subsequent developments. It is useful to record the relations $\omega_0 t_{\rm rr} = \pi/\epsilon^2$ and $\omega_0 t_{\rm res} = \pi/\epsilon$.

While $\tau$ is a convenient time variable to describe the entirety of an inspiral, the transit through a resonance is best described in terms of the rescaled variable $\lambda := (t-t_0)/t_{\rm res} = \tau/\epsilon$. Over a time interval around $t_0$ that is much shorter than $t_{\rm rr}$, we have that Eqs.~(\ref{p_omega}) and (\ref{orb_phase}) are usefully approximated by
\begin{equation}
p = p_0 \bigl[ 1 + O(\epsilon \lambda) \bigr], \qquad
\omega = \omega_0\bigl[ 1 + O(\epsilon \lambda) \bigr], \qquad 
\Phi = \Phi_0 + \frac{\pi}{\epsilon} \lambda + \frac{\pi}{2} \lambda^2 + O(\epsilon \lambda^3).
\label{orbit_short}
\end{equation}
It is useful to note that $\pi \lambda /\epsilon = \omega_0(t-t_0)$.

\subsection{Mode amplitude across a resonance}
\label{subsec:across} 

We return to the mode equation (\ref{q_eq_inertial}),
\begin{equation} 
\dot{q}_n^m + i\omega_n^m q_n^m = i f_n^m, 
\end{equation} 
and the forcing term of Eq.~(\ref{f_nm_fixedorbit}), which we re-express as 
\begin{equation}
f_n^m = A^m_{+n} e^{-i\Phi} + A^m_{-n} e^{i\Phi}
\end{equation}
to account for the inspiral; the coefficients $A^m_{\pm n}$ are listed in Eq.~(\ref{Apm}), and they are now slowly-varying functions of time. The solution to the mode equation, assuming an unperturbed state in the remote past, is
\begin{equation}
q_n^m(t) = i e^{-i\omega_n^m t} \biggl\{
\int_{-\infty}^t A_{+n}^m(t') e^{i[\omega_n^m t' - \Phi(t')]}\, dt' 
+ \int_{-\infty}^t A_{-n}^m(t') e^{i[\omega_n^m t' + \Phi(t')]}\, dt' \biggr\}. 
\label{q_sol} 
\end{equation} 
We are interested in the development of a resonance, which occurs when $\omega(t')$ momentarily becomes equal to  $|\omega_n^m|$. When this happens, one of the exponential factors inside the integrals stops oscillating (the phase becomes stationary), and this allows the integral to grow significantly. When $\omega_n^m > 0$, only the first integral participates in the resonance; the second integral does not grow. Only the second integral participates when $\omega_n^m < 0$.  

We wish to construct the solution to the mode equation before, during, and after the development of a resonance. To keep the following discussion simple we shall examine the specific case of a positive-frequency mode; the calculations for a negative-frequency mode require very few alterations, and our final results will apply to both cases. We simplify the notation by setting $q_n^m = q$, $\omega_n^m = \omega_0 > 0$, and $A_{+n}^m = A$. Because the second integral is now irrelevant in Eq.~(\ref{q_sol}), the mode equation can be simplified to
\begin{equation} 
\dot{q} + i\omega_0 q = i A e^{-i\Phi}.
\label{modeeq_simple} 
\end{equation}

The solution to Eq.~(\ref{modeeq_simple}) at early times, before resonance, is
\begin{equation}
q(t) = \frac{A(t) e^{-i\Phi(t)}}{\omega_0 - \omega(t)} \biggl[ 1
  + O\biggl( \frac{1}{(\omega_0-\omega) t_{\rm rr}} \biggr) \biggr].
\label{q_early1} 
\end{equation}
This is obtained by writing down the exact solution as in Eq.~(\ref{q_sol}), decomposing $\Phi(t')$ into a rapidly-varying piece $\omega(t)(t'-t)$ and a slowly-varying piece $\Phi_{\rm slow}(t')$, and integrating by parts to generate derivatives of $A$ and $\Phi_{\rm slow}$; the error term in Eq.~(\ref{q_early1}) comes from these derivatives, if we note that $A$ and $\Phi_{\rm slow}$ vary over a time scale comparable to $t_{\rm rr}$.

We may examine Eq.~(\ref{q_early1}) at times $t$ such that $t_0 - t$ is small compared with $t_{\rm rr}$, but still large enough to keep the error term small. We switch to the variable $\lambda$ defined near the end of Sec.~\ref{subsec:inspiral}, and we expand $A(t)$ and $\Phi(t)$ in powers of $\epsilon \ll 1$, as in Eq.~(\ref{orbit_short}). We arrive at
\begin{equation}
q(t) = -\frac{A(t_0)}{\epsilon \omega_0 \lambda}
e^{-i[\Phi_0 + \frac{\pi}{\epsilon} \lambda + \frac{\pi}{2} \lambda^2]}
\Bigl[ 1 + O(\epsilon \lambda, \lambda/\epsilon) \Bigr].
\label{q_early2} 
\end{equation}
This provides an accurate description of the solution for times $\lambda$ such that $\epsilon \ll |\lambda| \ll \epsilon^{-1}$.

Next we construct a solution to Eq.~(\ref{modeeq_simple}) that is valid when $|\lambda| \ll \epsilon^{-1}$; this will describe the transit through the resonance. We once again expand $A$ and $\Phi$ in powers of $\epsilon \ll 1$, and rewrite the mode equation as
\begin{equation}
\frac{dq}{d\lambda} + i \frac{\pi}{\epsilon} q = i t_{\rm res} A(t_0) 
e^{-i[\Phi_0 + \frac{\pi}{\epsilon} \lambda + \frac{\pi}{2} \lambda^2]}.
\end{equation}
The solution to this equation that is compatible with Eq.~(\ref{q_early2}) at early times is
\begin{equation}
q(t) = i t_{\rm res} A(t_0) e^{-i[\Phi_0 + \frac{\pi}{\epsilon} \lambda]} \Bigl\{
\bigl[ C(\lambda) + \tfrac{1}{2} \bigr] - i \bigl[ S(\lambda) + \tfrac{1}{2} \bigr] \Bigr\},
\label{q_during}
\end{equation}
where $C(\lambda)$ and $S(\lambda)$ are the Fresnel integrals [see, for example, Chapter 7 of Ref.~\cite{NIST:10}] 
\begin{equation}
C(\lambda) := \int_0^\lambda \cos\Bigl( \frac{\pi}{2} \lambda^{\prime 2} \Bigr)\, d\lambda' ,
\qquad 
S(\lambda) := \int_0^\lambda \sin\Bigl( \frac{\pi}{2} \lambda^{\prime 2} \Bigr)\, d\lambda'. 
\end{equation}
That Eq.~(\ref{q_during}) reproduces the behavior required by Eq.~(\ref{q_early2}) is verified with the asymptotic relations 
\begin{equation}
C(\lambda) = \frac{1}{2} \mbox{sign}(\lambda)
+ \frac{1}{\pi\lambda} \sin\Bigl( \frac{\pi}{2} \lambda^2 \Bigr) + O(\lambda^{-3}),
\qquad
S(\lambda) = \frac{1}{2} \mbox{sign}(\lambda)
- \frac{1}{\pi\lambda} \cos\Bigl( \frac{\pi}{2} \lambda^2 \Bigr) + O(\lambda^{-3}).
\label{Fresnel_asymp} 
\end{equation}
In this context it is useful to recall that $t_{\rm res} = \pi/(\epsilon \omega_0)$.

Equations (\ref{q_during}) and (\ref{Fresnel_asymp}) further reveal that beyond the resonance, when $\lambda \gg 1$, the mode amplitude is well approximated by
\begin{equation}
q(t) = \sqrt{2} t_{\rm res} A(t_0) e^{-i[\Phi_0 + \frac{\pi}{\epsilon}\lambda - \frac{\pi}{4}]}
- \frac{t_{\rm res} A(t_0)}{\pi \lambda}
e^{-i[\Phi_0 + \frac{\pi}{\epsilon} \lambda + \frac{\pi}{2} \lambda^2]}. 
\label{q_late2} 
\end{equation}
The second term is the same as in Eq.~(\ref{q_early2}), and it represents a particular solution to the mode equation. The first term is a solution to the homogeneous version of the mode equation, and it describes a free oscillation at frequency $\pi/\epsilon$ in terms of $\lambda$, or $\omega_0$ in terms of $t$. This free oscillation is the result of a resonant excitation of the mode associated with the amplitude $q$.

Equation (\ref{q_early1}) is a valid solution to Eq.~(\ref{modeeq_simple}) after the resonance has occurred, just as it was prior to its occurrence. But this particular solution must be augmented by a solution to the homogeneous equation to properly account for the resonant excitation of the mode. The solution that matches the behavior required by Eq.~(\ref{q_late2}) is
\begin{equation}
q(t) = \sqrt{2} t_{\rm res} A(t_0) e^{-i[\Phi_0 + \omega_0(t-t_0) - \frac{\pi}{4}]} 
- \frac{A(t) e^{-i\Phi(t)}}{\omega(t) - \omega_0} \biggl[ 1
  + O\biggl( \frac{1}{(\omega-\omega_0) t_{\rm rr}} \biggr) \biggr].
\label{q_late1} 
\end{equation}
By patching together Eqs.~(\ref{q_early1}), (\ref{q_during}), and (\ref{q_late1}) we obtain a complete description of the mode amplitude before, during, and after the onset of resonance.

Restoring the original notation for the mode amplitudes, and allowing for the existence of negative-frequency modes, we have found that Eq.~(\ref{q_sol}) is well approximated by 
\begin{equation}
q_n^m(t) = i t_{\rm res} A_{\pm n}^m(t_n^m) e^{\mp i[\Phi_n^m + \frac{\pi}{\epsilon} \lambda]} \Bigl\{
\bigl[ C(\lambda) + \tfrac{1}{2} \bigr] \mp i \bigl[ S(\lambda) + \tfrac{1}{2} \bigr] \Bigr\}
\label{q_during_final}
\end{equation}
when $|\lambda| \ll \epsilon^{-1}$. Here, $\lambda := (t-t_n^m)/t_{\rm res}$ is the time variable introduced near the end of Sec.~\ref{subsec:inspiral}, scaled in terms of the resonance-transit time of Eq.~(\ref{tres_def}), in which we insert $\omega_0 = |\omega_n^m|$. In addition, $t_n^m$ (formally denoted $t_0$) is the time at which $\omega(t) = |\omega_n^m|$, and $\Phi_n^m$ (formally denoted $\Phi_0$) is the orbital phase at time $t = t_n^m$. Equation (\ref{q_during_final}) involves a choice of coefficients $A_{\pm n}^m$, as listed in Eq.~(\ref{Apm}), a choice of sign in the exponential factor, and another choice of sign within the curly brackets; in all cases we select the upper sign for a positive-frequency mode, and the lower sign for a negative-frequency mode. 

A consequence of Eq.~(\ref{q_during_final}) is that
\begin{equation}
\dot{q}_n^m = -i \omega_n^m q_n^m \bigl[ 1 + O(\epsilon) \bigr]
\label{qdot_vs_q}
\end{equation} 
when $|\lambda| \ll \epsilon^{-1}$. This relation will be put to good use in subsequent developments.

\subsection{Mode-orbit coupling}
\label{subsec:mode-orbit}

The growth of a mode during the development of a resonance has a dynamical impact on the orbit. In the case of a stable mode, the energy given to the mode is taken away from the orbital motion; this produces a faster inspiral. In the case of an $r$-mode subjected to the Chandrasekhar-Friedman-Schutz instability, the star's rotational energy is given in part to the mode and in part to the orbital motion; this produces a slower inspiral. The mode-orbit interaction is mediated by the star's current quadrupole moment $J^{jk}$, which produces a post-Newtonian perturbing force on the companion. In this subsection we describe the effect of this force on the orbital motion.

We re-introduce the orbit-based vectorial basis of Sec.~\ref{sec:gravitomagnetic}, suitably modified to account for the inspiral. We have $\bm{n} = [\cos\Phi, \cos\iota\, \sin\Phi, \sin\iota\, \sin\Phi]$ pointing in the direction of the companion, $\bm{\lambda} = [-\sin\Phi, \cos\iota\, \cos\Phi, \sin\iota\, \sin\Phi]$ pointing in the direction of the orbital velocity, and $\bm{l} = \bm{n} \times \bm{\lambda} = [0, -\sin\iota\, \cos\iota]$ denoting the normal to the orbital plane.

When we account for a perturbing acceleration $\bm{a}$ in addition to the radiation-reaction force, the rate of change of the orbital radius is modified from Eq.~(\ref{pdot}) and becomes [see, for example, Eq.~(3.64a) of Ref.~\cite{poisson-will:14}] 
\begin{equation}
\dot{p} = -\frac{64}{5} c^{-5} (G {\cal M})^{5/3} (G M_{\rm tot})^{4/3} p^{-3}
+ \frac{2}{\omega}\,  {\cal S},
\label{pdot_pert}
\end{equation}
where ${\cal S} := \bm{a} \cdot \bm{\lambda}$ is the tangential component of the perturbing acceleration. To integrate Eq.~(\ref{pdot_pert}) we write $p = p_{\rm ins} + \delta p$, where $p_{\rm ins}$ is the inspiral motion described by Eq.~(\ref{p_omega}), and $\delta p$ is the perturbation. Linearizing with respect to $\delta p$, the differential equation becomes
\begin{equation}
\delta \dot{p} = \frac{2}{t_{\rm rr}} (1 - \tfrac{8}{3} \tau)^{-1}\, \delta p
+ \frac{2}{\omega_0} (1 - \tfrac{8}{3} \tau)^{3/8}\, {\cal S},
\end{equation}
and its solution is
\begin{equation}
\delta p(t) = \frac{2}{\omega_0} (1 - \tfrac{8}{3} \tau)^{-3/4}
\int_{-\infty}^t (1 - \tfrac{8}{3} \tau')^{9/8}\, {\cal S}(t')\, dt'.
\label{deltap_formal}
\end{equation}
In this equation, $\omega_0$ is the reference frequency introduced in Sec.~\ref{subsec:inspiral}, which will eventually be identified with one of the resonant frequencies.

The integral in Eq.~(\ref{deltap_formal}) accumulates mostly during the resonance. After the resonance, $\delta p(t)$ is well approximated by letting the upper bound of integration go to infinity. The expression then implies that $p = p_{\rm ins} + \delta p$ again corresponds to an inspiral, but one that is shifted in time relative to the original description. We have that
\begin{equation}
p(t) = p_{\rm ins}(t) + \delta p(t) = p_{\rm ins}(t + \Delta t),
\label{p_timeshift} 
\end{equation}
where
\begin{equation}
\Delta t = -\frac{3 t_{\rm rr}}{p_0 \omega_0} 
\int_{-\infty}^\infty (1 - \tfrac{8}{3} \tau)^{9/8}\, {\cal S}(t)\, dt. 
\label{Deltat_def}
\end{equation}
The meaning of this result is that the perturbation has essentially no influence on the orbit before and after the resonance; in both cases the evolution of the orbital radius is described by Eq.~(\ref{p_omega}). But the resonance produces a sudden shift in orbital radius, which is manifested by the time shift $\Delta t$ in Eq.~(\ref{p_timeshift}). Another way of stating this is that the inspirals before and after the resonance are both described by a solution to Eq.~(\ref{pdot}), but that the solutions come with different boundary conditions.

A crude description of the transit through a resonance, justified on the grounds that $t_{\rm res} \ll t_{\rm rr}$, is one in which $p$ and $\omega$ are taken to change discontinuously at $t = t_0$, but are otherwise not affected by the mode-orbit coupling. We write 
\begin{equation}
p(t) = \left\{ \begin{array}{ll}
p_{\rm ins}(t) & \qquad t < t_0 \\
p_{\rm ins}(t + \Delta t) & \qquad t > t_0
\end{array} \right., 
\end{equation}
where $\Delta t$ is the time shift of Eq.~(\ref{Deltat_def}). According to this, the instantaneous jump in orbital radius at resonance is given by $\Delta t\, \dot{p}_{\rm ins}(t_0)$; a positive $\Delta t$ gives rise to a decrease in orbital radius. 

The orbital phase is taken to change continuously across the resonance, and for this we write 
\begin{equation}
\Phi(t) = \left\{ \begin{array}{ll}
\Phi_{\rm ins}(t) & \qquad t < t_0 \\
\Phi_{\rm ins}(t + \Delta t) - \Delta \Phi & \qquad t > t_0
\end{array} \right., 
\label{phase_model} 
\end{equation}
where 
\begin{equation}
\Delta \Phi = \omega_0 \Delta t
\label{Deltaphi_def}
\end{equation}
is a constant introduced to ensure that $\Phi(t)$ is indeed continuous at $t = t_0$. This is an important quantity: Flanagan and Racine \cite{flanagan-racine:07} showed that in a measurement of gravitational waves emitted by a compact binary, $2\Delta \Phi$ is the accumulated phase difference between a model signal that properly incorporates the resonance and another model that does not; the factor of two accounts for the fact that the gravitational-wave frequency is twice the orbital frequency. A sizable phase shift implies that resonance effects can be detected in the gravitational-wave signal.   

Equation Eq.~(\ref{Deltaphi_def}) was derived on the basis of our crude description of the transit through a resonance. In Appendix~\ref{sec:phaseshift} we show that it also follows from a more elaborate treatment. 

\subsection{Calculation of the orbital phase shifts} 

The perturbing acceleration $\bm{a}$ created by a body with current quadrupole moment $J^{jk}$ was obtained by Flanagan and Racine \cite{flanagan-racine:07}, as a special case of their post-Newtonian equations of motion for arbitrarily shaped bodies \cite{racine-flanagan:05}. 
It is given by
\begin{equation}
a^j = \frac{4G}{c^2} \frac{M_{\rm tot}}{M} \bigg( -\epsilon^j_{\ kp} \dot{J}^p_{\ q} \frac{n^\stf{kq}}{p^3}
+ 5 \epsilon^j_{\ kp} J^p_{\ q} v'_s \frac{n^\stf{kqs}}{p^4}
- 5 \epsilon^k_{\ pq} J^q_{\ s} v'_k \frac{n^\stf{jps}}{p^4} \biggr),
\end{equation}
where $v'_k$ are the components of the orbital velocity vector, and $n^\stf{jk} := n^j n^k - \frac{1}{3} \delta^{jk}$, $n^\stf{jkp} := n^j n^k n^p - \frac{1}{5}(\delta^{jk} n^p + \delta^{jp} n^k + \delta^{kp} n^j)$ are symmetric-tracefree combinations of the vector $n^j$. For a circular orbit we have that $v'_k = p\omega \lambda_k$, and we find that the second and third terms in $a^j$ cancel out when we calculate ${\cal S} = a^j \lambda_j$. We get
\begin{equation}
{\cal S} = \frac{4G}{c^2} \frac{M_{\rm tot}}{M} \frac{1}{p^3} \dot{J}_{jk} l^j n^k
\label{S_result}
\end{equation}
after simplification.

The current quadrupole moment was expressed in terms of the mode amplitudes $q_n^m$ in Eqs.~(\ref{Jjk_mode1}) and (\ref{Jjk_mode3}). We substitute this in Eq.~(\ref{S_result}), make use of
Eq.~(\ref{qdot_vs_q}) to approximately equate $\dot{q}_n^m$ to $-i\omega_n^m q_n^m$, and invoke   
Eq.~(\ref{p_omega}) to express $p \simeq p_{\rm ins}$ as a function of time. We obtain   
\begin{equation}
{\cal S} = \sum_{m\geq 0} \psum_n {\cal S}_n^m
\label{S_mode1} 
\end{equation}
with
\begin{equation} 
{\cal S}_n^m = \frac{2R^3}{c^2} (m\Omega - \omega_n^m) (\omega_n^m)^3 \gothp_n^m
(1 - \tfrac{8}{3} \tau)^{-3/4} \bigl[ a_+^m \mbox{Re}(i q_n^m e^{i\Phi})
+ a_-^m \mbox{Re}(i q_n^m e^{-i\Phi}) \bigr],
\label{S_mode2}
\end{equation}
where
\begin{equation}
a_\pm^2 = \mp \sin\iota (\cos\iota \pm 1), \qquad
a_\pm^1 = \pm (\cos\iota \pm 1)(2\cos\iota \mp 1), \qquad
a_\pm^0 = \mp \sin\iota \cos\iota.
\label{a_pm}
\end{equation}
In Eq.~(\ref{S_mode2}), $\tau := (t-t_n^m)/t_{\rm rr}$ is the time variable introduced in Sec.~\ref{subsec:inspiral}, refined so that it refers specifically to the resonance time $t_n^m$ at which $\omega(t) = |\omega_n^m|$; the radiation-reaction time $t_{\rm rr}$ is defined by Eq.~(\ref{trr_def}), with $\omega_0 = |\omega_n^m|$. We recall that the prime on the sum over $n$ in Eq.~(\ref{S_mode1}) is an instruction for $m = 0$ to include positive-frequency modes only.

To calculate the time shift $(\Delta t)_n^m$ for a mode labelled by $m$ and $n$, we substitute  Eq.~(\ref{q_during_final}) for $q_n^m$ within Eq.~(\ref{S_mode2}), which we then insert within Eq.~(\ref{Deltat_def}). For a positive-frequency mode, only the term involving $i q_n^m e^{i\Phi}$ in Eq.~(\ref{S_mode2}) produces a significant contribution to the integral; for a negative-frequency mode, the only relevant term is the one involving $i q_n^m e^{-i\Phi}$. In either case we obtain
\begin{align}
(\Delta t)_n^m &= \pm \frac{6R^3}{c^2} (G M_{\rm tot})^{-1/3} t_{\rm rr} t_{\rm res}^2 a_\pm^m
A_{\pm n}^m(t_n^m) \gothp_n^m (m\Omega - \omega_n^m) |\omega_n^m|^{8/3}
\nonumber \\ & \quad \mbox{} \times 
\int_{-\infty}^\infty (1 - \tfrac{8}{3} \tau)^{3/8} \Bigl\{
\bigl[ C(\lambda) + \tfrac{1}{2} \bigr] \cos\frac{\pi}{2} \lambda^2
+ \bigl[ S(\lambda) + \tfrac{1}{2} \bigr] \sin\frac{\pi}{2} \lambda^2 \Bigr\}\, d\lambda.
\label{Deltat_intermediate} 
\end{align}
To evaluate the integral we observe that the factor within curly brackets is very small\footnote{It behaves as $-\pi^{-2} \lambda^{-3}$.} when $\lambda \ll -1$, increases markedly in the interval $|\lambda| \lesssim 1$, and oscillates increasingly rapidly when $\lambda \gg 1$. The integral is therefore dominated by the second interval, in which we can set $(1 - \frac{8}{3} \tau)^{3/8} = (1 - \frac{8}{3} \epsilon \lambda)^{3/8} = 1 + O(\epsilon)$. Evaluation of the integral is then straightforward, and we obtain a value of unity.

The time shift for a mode labelled by $m$ and $n$ is therefore given by
\begin{equation}
(\Delta t)_n^m = \pm \frac{6R^3}{c^2} (G M_{\rm tot})^{-1/3} t_{\rm rr} t_{\rm res}^2 a_\pm^m
A_{\pm n}^m(t_n^m) \gothp_n^m (m\Omega - \omega_n^m) |\omega_n^m|^{8/3}. 
\end{equation}
We next incorporate Eq.~(\ref{trr_def}) for $t_{\rm rr}$ and (\ref{tres_def}) for $t_{\rm res}$, replacing $\omega_0$ in these expressions with $|\omega_n^m|$. We also insert Eqs.~(\ref{Apm}) for $A_{\pm n}^m$, Eq.~(\ref{Bjk}) for ${\cal B}_0$, and Eq.~(\ref{chirp_mass}) for the chirp mass ${\cal M}$. Writing $\omega_n^m = w_n^m \Omega$ for the mode frequencies, we finally arrive at
\begin{equation}
(\Delta t)_n^m = \frac{25\pi^2}{2304} \frac{(\gothp_n^m)^2}{\hat{N}_n^m}
\frac{(m-w_n^m)^2}{|w_n^m|^{7/3}} (a_\pm^m)^2 
\frac{c^6 R^4}{(G M)^2 (GM') (GM_{\rm tot})^{1/3} \Omega^{1/3}}.
\label{Deltat_final} 
\end{equation}

The orbital phase shift $(\Delta \Phi)_n^m$ associated with a mode labelled by $m$ and $n$ is obtained by inserting Eq.~(\ref{Deltat_final}) within Eq.~(\ref{Deltaphi_def}), once again replacing $\omega_0$ with $|\omega_n^m|$. We get
\begin{equation}
(\Delta \Phi)_n^m = \frac{25\pi^2}{2304} \frac{(\gothp_n^m)^2}{\hat{N}_n^m}
\frac{(m-w_n^m)^2}{|w_n^m|^{4/3}} (a_\pm^m)^2 
\frac{c^6 R^4 \Omega^{2/3}}{(G M)^2 (GM') (GM_{\rm tot})^{1/3}}.
\label{Deltaphi_final} 
\end{equation}
We recall that the overlap integrals $\gothp_n^m$ were defined in Eq.~(\ref{overlap2}), and that the mode norms $\hat{N}_n^m$ were introduced in Eq.~(\ref{Nhat_def}); these constants are displayed in Table~\ref{tab:overlaps} for various stellar models. In addition, we recall that the constants $a_\pm^m$ are listed in Eq.~(\ref{a_pm}).  

\subsection{Gravitational-wave phase shift} 

As the inspiral proceeds, the binary goes through a succession of four gravitomagnetic tidal resonances, one for each of the dominant Lockitch-Friedman inertial modes (including the $r$-mode). At each resonance the orbital phase undergoes a shift given by Eq.~(\ref{Deltaphi_final}), and the total accumulated phase shift in the gravitational-wave signal is 
\begin{equation} 
\Delta\Phi_{\rm GW} = \gamma(\iota) \biggl( \frac{R}{10\ \mbox{km}} \biggr)^4 
\biggl( \frac{1.4\ M_\odot}{M} \biggr)^2 
\biggl( \frac{1.4\ M_\odot}{M'} \biggr) 
\biggl( \frac{2.8\ M_\odot}{M_{\rm tot}} \biggr)^{1/3} 
\biggl( \frac{\Omega/2\pi}{100\ \mbox{Hz}} \biggr)^{2/3}, 
\label{DeltaPhi_GW1} 
\end{equation} 
where
\begin{equation}
\gamma(\iota) := \gothu_\bullet^2\, \sin^2\iota(\cos\iota + 1)^2 
+ \gothu_{\rm I}^1\, (\cos\iota + 1)^2 (2\cos\iota - 1)^2 
+ \gothu_{\rm II}^1\, (\cos\iota - 1)^2 (2\cos\iota + 1)^2
+ \gothu_{\rm I}^0\, \sin^2\iota \cos^2\iota
\label{DeltaPhi_GW2} 
\end{equation} 
with 
\begin{equation} 
\gothu_n^m := 2 \times 11.550 \times \frac{25\pi^2}{2304} \frac{(\gothp_n^m)^2}{\hat{N}_n^m}
\frac{(m-w_n^m)^2}{|w_n^m|^{4/3}}. 
\label{unm_def} 
\end{equation} 
In this expression, the factor of 2 accounts for the fact that the gravitational-wave frequency is twice the orbital frequency, the numerical factor of $11.550$ corresponds to the choice of fiducial values for $R$, $M$, $M'$, and $\Omega$, and the remaining factors come from Eq.~(\ref{Deltaphi_final}). 

Equations (\ref{DeltaPhi_GW1}) and (\ref{DeltaPhi_GW2}) as the same as Eqs.~(\ref{DeltaPhi_intro}) and (\ref{gamma_intro}), respectively. The numbers $\gothu_n^m$ are tabulated for selected values of the density-model parameter $k$ in Table~\ref{tab:uvalues_intro}, and plots of $\gamma$ as a function of the inclination angle $\iota$ are displayed in Fig.~\ref{fig:phaseshift_intro}. An in-depth discussion of these results was provided back in Sec.~\ref{sec:intro}, and there is no need to repeat it here. 

\begin{acknowledgments} 
I am grateful to John Friedman and Huan Yang for essential discussions that shaped the development of this work. I also thank Simon Alexandre Pekar, who verified some of my numerical results with independent computations, and Eanna Flanagan for helpful comments on an earlier draft of the paper. This work was supported by the Natural Sciences and Engineering Research Council of Canada.  
\end{acknowledgments} 

\appendix 

\section{Orbital phase through a resonance} 
\label{sec:phaseshift} 

In Sec.~\ref{subsec:mode-orbit}, following Flanagan and Racine \cite{flanagan-racine:07}, we derived the relation 
\begin{equation}
\Delta \Phi = \omega_0 \Delta t
\label{Deltaphi_repeat}
\end{equation}
on the basis of a crude resonance model in which $p$ and $\omega$ jump discontinuously at $t=t_0$, while $\Phi$ changes continuously. Here we provide a more credible derivation that does not rely on the pretence of an instantaneous resonance.  

The orbital phase $\Phi$ is governed by the differential equation $\dot{\Phi} = \omega = (GM_{\rm tot}) p^{-3/2}$. If we write $p = p_{\rm ins} + \delta p$ and $\Phi = \Phi_{\rm ins} + \delta \Phi$, where $p_{\rm ins}$ is given by Eq.~(\ref{p_omega}) and $\Phi_{\rm ins}$ by Eq.~(\ref{orb_phase}), then we find that $\delta \Phi$ satisfies
\begin{equation} 
\delta \dot{\Phi} = -\frac{3 \omega_0}{2 p_0} (1 - \tfrac{8}{3} \tau)^{-5/8}\, \delta p. 
\label{Phidot} 
\end{equation} 
The perturbation $\delta p$ is given by Eq.~(\ref{deltap_formal}), and taking Eq.~(\ref{Deltat_def}) into account, this is 
\begin{equation}
\delta p(t) = -\frac{2 p_0 \Delta t}{3 t_{\rm rr}} (1 - \tfrac{8}{3} \tau)^{-3/4}
\frac{\int_{-\infty}^t (1 - \tfrac{8}{3} \tau')^{9/8}\, {\cal S}(t')\, dt'} 
{\int_{-\infty}^\infty (1 - \tfrac{8}{3} \tau)^{9/8}\, {\cal S}(t)\, dt}. 
\end{equation} 
Incorporating Eq.~(\ref{S_mode1}) and retracing the steps that led to Eq.~(\ref{Deltat_intermediate}), we obtain 
\begin{equation}
\delta p(t) = -\frac{2 p_0 \Delta t}{3 t_{\rm rr}} (1 - \tfrac{8}{3} \tau)^{-3/4}\, {\cal F}(\lambda), 
\label{deltap_vs_F} 
\end{equation} 
where 
\begin{equation} 
{\cal F}(\lambda) = 
\frac{\int_{-\infty}^\lambda (1 - \tfrac{8}{3} \tau')^{3/8} \Bigl\{
\bigl[ C(\lambda') + \tfrac{1}{2} \bigr] \cos\frac{\pi}{2} \lambda^{\prime 2}
+ \bigl[ S(\lambda') + \tfrac{1}{2} \bigr] \sin\frac{\pi}{2} \lambda^{\prime 2} \Bigr\}\, d\lambda'} 
{\int_{-\infty}^\infty(1 - \tfrac{8}{3} \tau)^{3/8} \Bigl\{
\bigl[ C(\lambda) + \tfrac{1}{2} \bigr] \cos\frac{\pi}{2} \lambda^{2}
+ \bigl[ S(\lambda) + \tfrac{1}{2} \bigr] \sin\frac{\pi}{2} \lambda^{2} \Bigr\}\, d\lambda}. 
\end{equation} 
To arrive at this we eliminated the sum over modes in Eq.~(\ref{S_mode1}), to better focus our attention on a single resonance at frequency $\omega_0$. Setting $(1 - \tfrac{8}{3} \tau')^{3/8} = (1 - \tfrac{8}{3} \epsilon\lambda')^{3/8} = 1 + O(\epsilon)$ inside the integrals, we find that ${\cal F}(\lambda)$ is well approximated by 
\begin{equation} 
{\cal F}(\lambda) = \frac{1}{2} C(C+1) + \frac{1}{2} S(S+1) + \frac{1}{4}. 
\end{equation} 
Making the substitution in Eq.~(\ref{deltap_vs_F}), we obtain an explicit expression for $\delta p$ through the resonance. 

The solution to Eq.~(\ref{Phidot}) is 
\begin{equation}
\delta\Phi = \epsilon \omega_0 \Delta t \int_{-\infty}^\lambda 
(1 - \tfrac{8}{3} \tau')^{-11/8} {\cal F}(\lambda')\, d\lambda', 
\end{equation} 
where we used the facts that $dt/d\lambda = t_{\rm res} = \epsilon t_{\rm rr}$. Setting $(1 - \tfrac{8}{3} \tau')^{-11/8} = 1 + O(\epsilon)$ as usual, we use 
\begin{equation} 
\int_{-\infty}^\lambda {\cal F}(\lambda')\, d\lambda' = \lambda {\cal F} 
- \frac{1}{\pi} (C + \tfrac{1}{2}) \sin\Bigl( \frac{\pi}{2} \lambda^2 \Bigr)
+ \frac{1}{\pi} (S + \tfrac{1}{2}) \cos\Bigl( \frac{\pi}{2} \lambda^2 \Bigr) 
= \lambda + O(\lambda^{-1}) 
\end{equation} 
and conclude that 
\begin{equation} 
\delta \Phi \simeq \epsilon \omega_0 \Delta t\, \lambda 
\label{deltaPhi1} 
\end{equation} 
when $\lambda \gg 1$. 

On the other hand, we saw back in Sec.~\ref{subsec:mode-orbit} that after the resonance, the orbital frequency is well approximated by a time-shifted inspiral, $\omega(t) = \omega_{\rm ins}(t + \Delta t)$. Integration gives $\Phi(t) = \Phi_{\rm ins}(t + \Delta t) - \Delta\Phi$ for the orbital phase, where $\Delta\Phi$ is a constant of integration. This means that $\delta\Phi = \omega_{\rm ins}(t) \Delta t - \Delta\Phi$, and incorporating the approximation $\omega_{\rm ins}(t) = \omega_0[1 + \epsilon\lambda + O(\epsilon^2\lambda^2)]$ obtained in Sec.~\ref{subsec:inspiral}, we arrive at 
\begin{equation} 
\delta\Phi \simeq \omega_0 \Delta t - \Delta\Phi + \epsilon \omega_0 \Delta t\, \lambda. 
\end{equation} 
Comparing this with Eq.~(\ref{deltaPhi1}), we conclude that $\Delta\Phi = \omega_0 \Delta t$. This is the statement of  
Eq.~(\ref{Deltaphi_repeat}). 

\bibliography{/Users/poisson/writing/papers/tex/bib/master}

\end{document}